\documentclass[a4paper,fleqn]{cas-dc}

\usepackage[numbers]{natbib}

\usepackage[per-mode=symbol]{siunitx}
\usepackage[acronyms]{glossaries}
\usepackage{xparse}
\usepackage{mathtools}
\usepackage{bm}
\usepackage{algorithm}
\usepackage{algpseudocode}
\usepackage{pifont}
\usepackage{subcaption}
\glsdisablehyper
\newacronym{IMU}{IMU}{inertial measurement unit}
\newacronym{IOE}{IOE}{inertial orientation estimation}
\newacronym{ENU}{ENU}{east-north-up}
\newacronym{OMC}{OMC}{optical motion capture}
\newacronym{BROAD}{BROAD}{Berlin Robust Orientation Estimation Assessment Dataset}
\newacronym{TAGP}{TAGP}{trial-agnostic generalized performance}
\newacronym{RMSE}{RMSE}{root mean square error}
\newacronym{IIR}{IIR}{infinite impulse response}
\newcommand{\dps}{\degree\per\second}
\newcommand{\mpss}{\meter\per\square\second}
\definecolor{darkblue}{rgb}{0,0,.6}
\hypersetup{colorlinks=true, breaklinks=true, linkcolor=darkblue, menucolor=darkblue, urlcolor=darkblue, citecolor=darkblue}

\def\equationautorefname~#1\null{(#1)\null}
\newcommand{\nvec}[3][]{\ensuremath{\coordtrafobrackets{#1}{\mathbf{#2}_{\mathrm{#3}}}}}
\newcommand{\nquat}[2][]{\nvec[{#1}]{q}{#2}}
\newcommand{\quat}[3][]{\ensuremath{\coordtrafobrackets{#1}{\prescript{#2}{#3}{\mathbf{q}}}}}
\newcommand{\quataa}[2]{\ensuremath{\left(#1\,@\,#2\right)}}
\newcommand{\coordtrafobrackets}[2]{\ifthenelse{\equal{#1}{}}{#2}{\left[#2\right]_{#1}}}
\newcommand{\rot}[2]{\operatorname{rot}\left(#1, #2\right)}
\newcommand{\tk}[1][]{\ensuremath{t_{k#1}}}
\newcommand{\Ts}[1][]{\ensuremath{T_\mathrm{s}}}
\newcommand{\gyr}[2][]{\ensuremath{\nvec[#1]{\bm{\omega}}{#2}}}
\newcommand{\tauAcc}{\ensuremath{\tau_\mathrm{acc}}}
\newcommand{\tauMag}{\ensuremath{\tau_\mathrm{mag}}}
\newcommand{\tagpx}{TAGP$_\mathrm{x}$}
\newcommand{\tps}{\ensuremath{^\intercal}}

\newcommand{\norm}[1]{\ensuremath{\left\lVert#1\right\rVert}}
\newcommand{\normalized}[1]{\ensuremath{\frac{#1}{\norm{#1}}}}
\DeclareMathOperator{\atantwo}{atan2}
\newcommand{\quatinv}{\ensuremath{^{-1}}}
\newcommand{\quatmult}{\ensuremath{\otimes}}

\newcommand{\imu}[1][]{\ensuremath{\mathcal{S}\mathrm{#1}}}
\newcommand{\earth}[1][]{\ensuremath{\mathcal{E}\mathrm{#1}}}
\newcommand{\earthtrue}{\ensuremath{\earth_\mathrm{true}}}
\newcommand{\inertiali}{\ensuremath{\mathcal{I}_i}}
\newcommand{\imui}{\ensuremath{\imu_i}}
\newcommand{\earthi}{\ensuremath{\earth_i}}

\newcommand{\quatimuimu}[3][]{\quat[#1]{\imu{#2}}{\imu{#3}}}
\newcommand{\quatimuearth}[3][]{\quat[#1]{\imu{#2}}{\earth{#3}}}

\newcommand{\cmark}{\ding{51}}
\newcommand{\xmark}{\ding{55}}
\newcommand{\sref}[2]{\hyperref[#2]{#1 \ref{#2}}}
\definecolor{C0}{HTML}{1F77B4}\definecolor{C0dark}{HTML}{195F90}\definecolor{C0dark2}{HTML}{103C5A}\definecolor{C0light}{HTML}{258FD8}\definecolor{C0light2}{HTML}{2CA9FF}
\definecolor{C1}{HTML}{FF7F0E}\definecolor{C1dark}{HTML}{CC660B}\definecolor{C1dark2}{HTML}{804007}\definecolor{C1light}{HTML}{FF7F0E}\definecolor{C1light2}{HTML}{FF7F0E}
\definecolor{C2}{HTML}{2CA02C}\definecolor{C2dark}{HTML}{238023}\definecolor{C2dark2}{HTML}{165016}\definecolor{C2light}{HTML}{35C035}\definecolor{C2light2}{HTML}{42F042}
\definecolor{C3}{HTML}{D62728}\definecolor{C3dark}{HTML}{AB1F20}\definecolor{C3dark2}{HTML}{6B1314}\definecolor{C3light}{HTML}{FF2E30}\definecolor{C3light2}{HTML}{FF2E30}
\ExplSyntaxOn

\NewDocumentCommand \inlinevec { s o m }
 {
  \IfBooleanTF {#1}
   { \vectaux*{#3} }
   { \IfValueTF {#2} { \vectaux[#2]{#3} } { \vectaux{#3} } }
  \tps
 }

\NewDocumentCommand \inlinerowvec { s o m }
 {
  \IfBooleanTF {#1}
   { \vectaux*{#3} }
   { \IfValueTF {#2} { \vectaux[#2]{#3} } { \vectaux{#3} } }
 }

\DeclarePairedDelimiterX \vectaux [1] {\lbrack} {\rbrack}
 { \, \dbacc_vect:n { #1 } \, }

\cs_new_protected:Npn \dbacc_vect:n #1
 {
  \seq_set_split:Nnn \l_tmpa_seq { , } { #1 }
  \seq_use:Nn \l_tmpa_seq { \enspace }
 }
\ExplSyntaxOff

\ExplSyntaxOn
\RenewDocumentCommand \emailauthor { m m }
   {
     \int_gincr:N \g_ead_int
     \seq_gput_right:Nn \g_stm_ead_seq
       {
         { \ttfamily \tl_to_str:n { #1 } }
       }
     }
\RenewDocumentCommand \printorcid { }
{
}
\ExplSyntaxOff

\begin{document}
\let\WriteBookmarks\relax
\def\floatpagepagefraction{1}
\def\dbltopfraction{1}
\def\textpagefraction{.001}

\shorttitle{VQF: Highly Accurate IMU Orientation Estimation with Bias Estimation and Magnetic Disturbance Rejection}

\shortauthors{Laidig and Seel}

\title [mode = title]{VQF: Highly Accurate IMU Orientation Estimation with Bias Estimation and Magnetic Disturbance Rejection}

\tnotemark[1]
\tnotetext[1]{\textbf{Accepted manuscript. Please cite the following version:} D. Laidig and T. Seel. ``VQF: Highly Accurate IMU Orientation Estimation with Bias Estimation and Magnetic Disturbance Rejection.'' Information Fusion 2023, 91, 187--204. \href{https://doi.org/10.1016/j.inffus.2022.10.014}{https://doi.org/10.1016/j.inffus.2022.10.014}.}

\author[1]{Daniel Laidig}[]

\cormark[1]

\ead{laidig@control.tu-berlin.de}

\credit{Conceptualization, Methodology, Software, Validation, Formal analysis, Investigation, Data curation, Writing -- original draft, Writing -- review \& editing, Visualization}

\affiliation[1]{
  organization={Control Systems Group, Technische Universität Berlin},
  postcode={10623},
  postcodesep={},
  city={Berlin},
  country={Germany}
}

\author[2]{Thomas Seel}[]

\ead{thomas.seel@fau.de}

\credit{Conceptualization, Methodology, Formal analysis, Writing - review \& editing, Supervision}

\affiliation[2]{
  organization={Department Artificial Intelligence in Biomedical Engineering, Friedrich-Alexander-Universität Erlangen-Nürnberg},
  postcode={91052},
  postcodesep={},
  city={Erlangen},
  country={Germany}
}

\cortext[1]{Corresponding author}

\ExplSyntaxOn
\cs_gset:Npn \__first_footerline:
{
  \group_begin:
  \small
  \sffamily
  \ifnum\theblind>0\relax
  \else
	\__short_authors:
  \fi
  \group_end:
}
\ExplSyntaxOff

\begin{abstract}
The miniaturization of MEMS-based inertial measurement units (IMUs) facilitates their widespread use in a growing number of application domains.
The fundamental sensor fusion task of orientation estimation is a prerequisite for most further data processing steps in inertial motion tracking, such as position and velocity estimation, joint angle estimation, and 3D visualization.
Errors in the estimated orientations severely affect all further processing steps.
Recent systematic comparisons of existing algorithms show that out-of-the-box accuracy is often low and that application-specific tuning is required to obtain high accuracy.
In the present work, we propose and extensively evaluate a quaternion-based orientation estimation algorithm that is based on a novel approach of filtering the acceleration measurements in an almost-inertial frame and that includes extensions for gyroscope bias estimation and magnetic disturbance rejection, as well as a variant for offline data processing.
In contrast to all existing work, we perform an extensive evaluation, using a large collection of publicly available datasets and eight literature methods for comparison.
The proposed method consistently outperforms all eight literature methods and achieves an average RMSE of \ang{2.9}, while the errors obtained with literature methods range from \ang{5.3} to \ang{16.7}.
This improved accuracy with respect to the state of the art is observed not only in average but also for each of several different motion characteristics, as well as for gyroscope bias estimation.
Since the evaluation was performed with one single fixed parametrization across a very diverse dataset collection, we conclude that the proposed method provides unprecedented out-of-the-box performance for a broad range of motions, sensor hardware, and environmental conditions.
This gain in orientation estimation accuracy is expected to advance the field of IMU-based motion analysis and provide performance benefits in numerous applications.
The provided open-source implementation makes it easy to employ the proposed method.
\end{abstract}


\maketitle

\section{Introduction}\label{sec:vqf_intro}

In recent years, MEMS-based \glspl{IMU} have become small, lightweight, and affordable.
New application domains in which they are used include, for example, sports \cite{mcginnis2012highly}, gait analysis \cite{laidig2021calibrationfree,wang2015stancephase}, rehabilitation monitoring \cite{nguyen2018using}, rehabilitation robotics \cite{passon2020inertialrobotic}, autonomous vehicles \cite{rodrigomarco2020multimodal}, aerial vehicles \cite{chao2010autopilots}, and kites \cite{freter2020motion}.
In all these applications, \glspl{IMU} are used to estimate variables of motion, such as orientations, velocities, and positions, either in real time or via postprocessing of recorded data.

\begin{figure*}[thpb]
  \centering
  \includegraphics{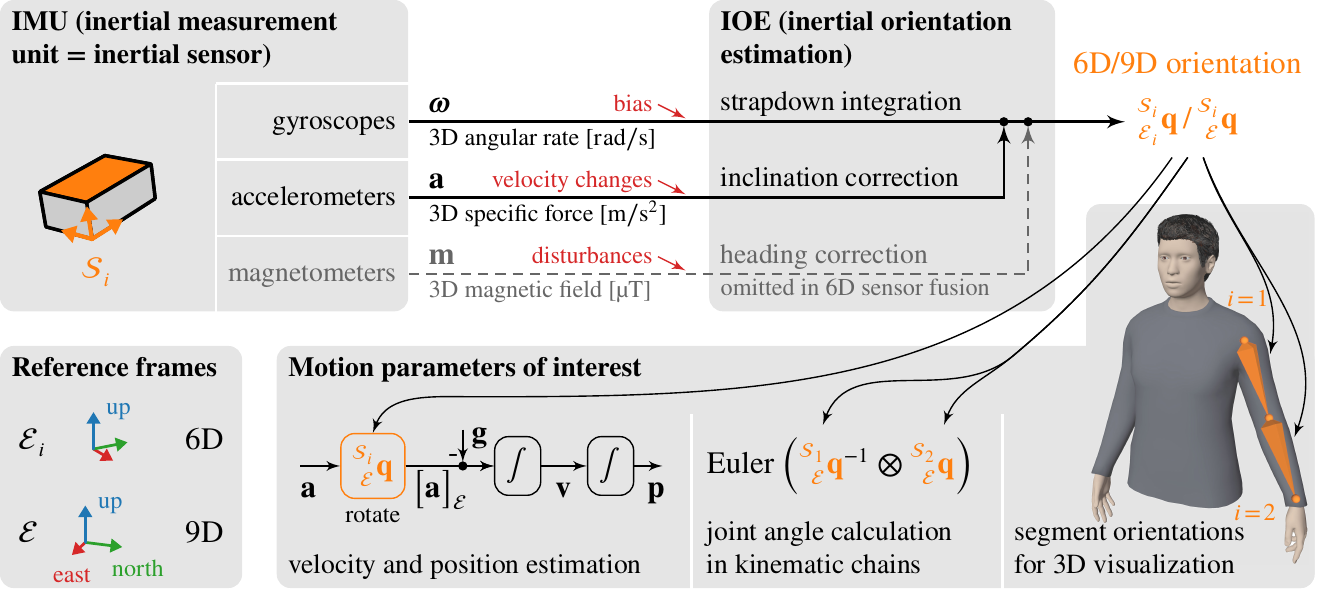}
  \caption{Orientation estimation of IMUs is achieved by sensor fusion of the gyroscope measurements with the accelerometer and, in 9D sensor fusion, magnetometer measurements.
  Obtaining an accurate orientation estimate is the prerequisite for fundamental further steps in inertial motion tracking, including velocity and position estimation, joint angle calculation, and 3D visualization.
  }
  \label{fig:vqf_introduction}
\end{figure*}

\glspl{IMU} measure angular rate, specific force (also called proper acceleration), and magnetic field strength, each as a time-dependent 3D vector in an intrinsic sensor coordinate system.
Those measurements are processed to determine the motion parameters of interest, e.g., the orientation of an object to which the sensor is attached, the object's velocity or position, or other application-specific motion parameters \cite{woodman2007introduction,kok2017using}.
As shown in \autoref{fig:vqf_introduction}, determining such motion parameters generally requires the prior estimation of the orientation of the sensor with respect to an inertial frame of reference, a procedure known as \emph{\gls{IOE}}.

Since \gls{IOE} is such a fundamental step in IMU-based motion analysis and the accuracy of all further parameters of interest depends on the accuracy of the orientation estimate, it is not surprising that abundant prior research has aimed at solving this task.
Comprehensive reviews that classify and compare the existing solution approaches are found in \cite{nazarahari202040,caruso2021analysis}.
The existing methods widely vary with respect to the filter type, the computational complexity, the number of tuning parameters, and the additional features such as gyroscope bias estimation.
For several algorithms, implementations in C++ or Matlab are available, see e.g. \cite{caruso2021analysis,valenti2015keeping,seel2017eliminating,kok2019fast,mahony2008nonlinear,madgwick2010efficient}.

In literature that proposes new \gls{IOE} algorithms, accuracy is most commonly validated using marker-based \gls{OMC} as a ground truth.
However, the employed datasets are almost always non-public and highly application-specific; they vary widely in terms of movement speed, characteristics of the employed motion, magnetic environment, and sensor error characteristics.
Therefore, reported performance figures cannot be compared directly.
This lack of common datasets and suitable benchmarks has recently been addressed by the publication of the Sassari dataset \cite{caruso2021orientation} and the \gls{BROAD} \cite{laidig2021broad}.
Newly proposed algorithms should be validated using such publicly available datasets in comparison with other state-of-the-art algorithms \cite{nazarahari202040}.

The few existing comparative studies show that the out-of-the-box performance of most algorithms is poor and that application-specific algorithm selection, as well as laborious parameter tuning, are necessary to achieve good results \cite{caruso2021analysis,laidig2021broad}, which represents a severe limitation of the state of the art.
Even with optimized parameters, the root-mean-square errors achieved by the best \gls{IOE} algorithms are in the range of \SIrange{1}{3}{\degree} for slow and smooth motions and as much as \SIrange{5}{15}{\degree} for fast and challenging motions \cite{caruso2021analysis,laidig2021broad}.
Further improving this accuracy seems highly desirable in view of numerous applications.

In summary, while there is ample work on various \gls{IOE} algorithms, evaluation of the proposed methods is often limited and cannot be compared across publications.
Recent comparative reviews and benchmarks show that there is no one-size-fits-all solution that works out of the box and yields high accuracy for a wide variety of application scenarios.
Furthermore, the widespread adoption of novel \gls{IOE} algorithms is not only driven by accuracy but also depends on the availability of an easy-to-use implementation.
In combination, this demonstrates that there is a need for an algorithm that is validated on a very large and diverse set of experimental data, provides accurate out-of-the-box orientation estimates without tuning, and is easy to use and to integrate into existing code projects.

We aim at filling this gap in two steps:
We first propose a new feature-rich quaternion-based orientation estimation algorithm and then perform an extensive validation to demonstrate the exceptionally high accuracy that is achieved by this algorithm.
With respect to the first step, the key differences of the proposed algorithm with respect to the latest state of the art are best expressed by the following five features:

\begin{enumerate}
\item As a novel approach to sensor fusion of gyroscopes and accelerometers, the accelerometer information is low-pass filtered in an almost-inertial frame, which yields robust rejection of accelerations due to velocity changes.
\item Magnetometer-based heading correction is performed as a modular decoupled step, which eliminates the influence of magnetic disturbances on the inclination and facilitates simultaneous 6D and 9D estimation.
\item The algorithm includes extensions for online gyroscope bias estimation during rest and motion and an optional magnetic disturbance rejection strategy.
\item In contrast to the vast majority of previous approaches, an acausal offline version is available, which further increases the accuracy in situations in which real-time capability is not required.
\item Easy-to-use open-source implementations of the proposed algorithms are provided in C++, Python, and Matlab.
\end{enumerate}
With respect to the second step, the main contributions and results of the extensive accuracy evaluation are:
\begin{enumerate}
\item In contrast to most previous work, the proposed method is extensively evaluated using a large collection of publicly available data and in comparison with eight existing \gls{IOE} algorithms.
\item The results show that the proposed method outperforms all evaluated existing methods, providing a 1.8-fold to 5-fold increase in orientation estimation accuracy.
\item For a large variety of motions, speeds, and disturbed environments, the proposed method works out of the box, and application-specific parameter tuning is not necessary.
\end{enumerate}

\section{Proposed Method for Inertial Orientation Estimation}\label{sec:vqf_method}

We briefly explain the employed terminology and notation and then, building upon preliminary work in \cite{laidig2022vqffusion}, propose a modular method for simultaneous 6D and 9D orientation estimation.

\subsection{Terminology and Notation}

As illustrated in \autoref{fig:vqf_introduction}, the following measurements are available in \gls{IOE}: gyroscope readings $\gyr{}(\tk)\in\mathbb{R}^3$, accelerometer readings $\nvec{a}{}(\tk)\in\mathbb{R}^3$, and magnetometer readings $\nvec{m}{}(\tk)\in\mathbb{R}^3$, sampled at times $\tk=k T_\mathrm{s}, k\in\{1, 2, \ldots, N\}, T_\mathrm{s}\in\mathbb{R}_{>0}$.

If only gyroscopes and accelerometers are employed, we use the term \emph{6D} \gls{IOE}, while \emph{9D} \gls{IOE} additionally uses magnetometers.
Therefore, 9D \gls{IOE} yields the sensor orientation with respect to a fixed inertial reference frame, typically using the \gls{ENU} convention (i.e., $z$ is pointing up and $y$ is pointing north).
In contrast, only vertical reference information is available in 6D \gls{IOE}, and the resulting orientations are thus provided with respect to an almost-inertial reference frame, which has one vertical axis and slowly drifts around this axis (at a rate determined by the gyroscope bias, i.e., typically $\le\SI{1}{\dps}$).

We denote the moving sensor frame, i.e., the coordinate system in which the sensor readings are provided, by $\imui(\tk)$.
The \gls{ENU} inertial reference frame used in 9D \gls{IOE} is denoted $\earth$, and the sensor-specific almost-inertial reference frame used in 6D \gls{IOE} is denoted $\earthi(\tk)$.
Anticipating the common application scenario with multiple \glspl{IMU} on different segments of a kinematic chain, a sensor index $i$ is used for sensor-specific quantities.

We use square brackets to specify the coordinate system in which a vector is expressed, for example, $\nvec[\earth]{a}{}$ is the accelerometer measurement transformed into frame \earth{}.
We denote rotations and orientations by unit quaternions \cite{kuipers1999quaternions} in vector notation, i.e., we write the quaternion $w + i x + j y + k z$ as $\inlinevec{w,x,y,z}$.
In the context of quaternion multiplication, which we denote by $\quatmult$, we implicitly regard 3D vectors as pure quaternions.
For example, $\nvec[\earth]{a}{} = \quatimuearth{}{} \quatmult \nvec{a}{} \quatmult \quatimuearth{}{} \quatinv$.
Here, the left upper and lower indices denote the frames between which the quaternion rotates.
Quaternions that represent the rotation of an angle $\alpha\in\mathbb{R}$ around the axis $\nvec{v}{}\in\mathbb{R}^3$ are written as
$\quataa{\alpha}{\nvec{v}{}} := \begin{bmatrix}\cos \frac{\alpha}{2} & \frac{\nvec{v}{}\tps}{\norm{\nvec{v}{}}}\sin \frac{\alpha}{2}\end{bmatrix}\tps$.

\begin{figure}[t]
\includegraphics{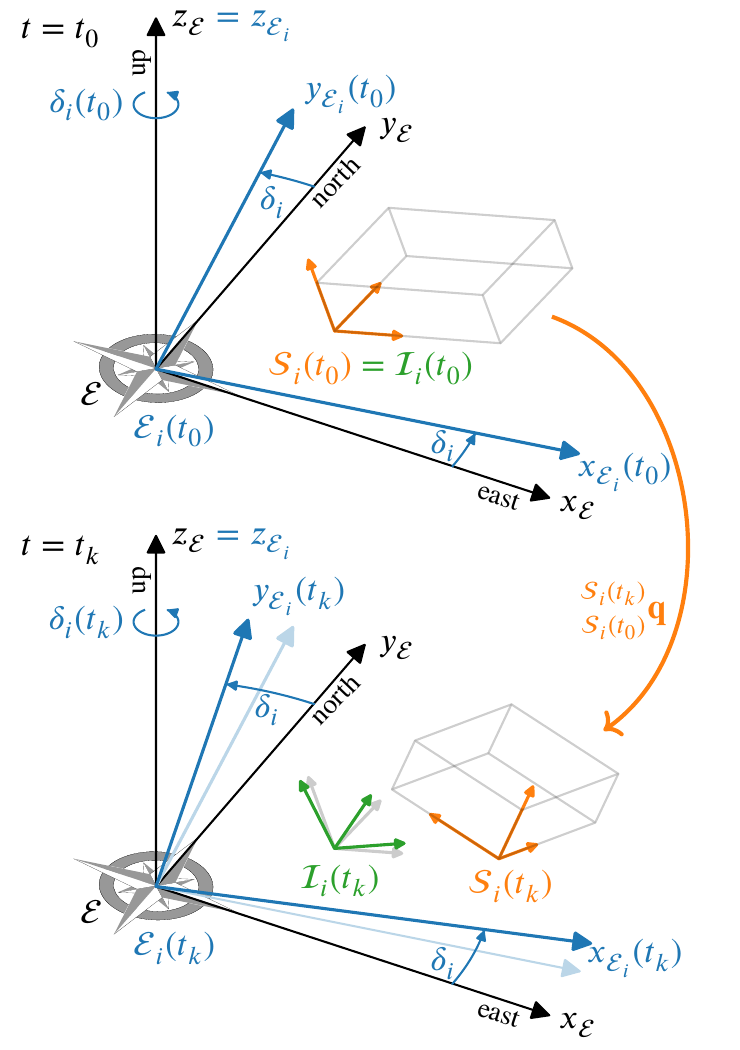}
\caption{Illustration of the different coordinate systems used by the proposed method. The aim of \gls{IOE} is to determine the orientation of the sensor $\imui$ relative to an \gls{ENU} reference frame $\earth$ (in 9D sensor fusion) or relative to a reference frame $\earth_i$ with vertical $z$-axis (in 6D sensor fusion). The angle $\delta_i$ describes the slowly drifting heading offset between $\earth$ and $\earthi$. Internally, the auxiliary $\inertiali$ frame is used to represent the orientation obtained by pure gyroscope strapdown integration and slowly drifts due to the integration of gyroscope bias.}
\label{fig:vqf_coordinate_systems}
\end{figure}

\subsection{A Modular Estimation Approach}

The most fundamental state of any \gls{IOE} algorithm is the current orientation estimate, which is commonly represented by a single orientation quaternion.
For the proposed method, we use a more modular approach and represent the 6D estimate $\quat{\imui(\tk)}{\earthi(\tk)}$ as the concatenation of an inclination correction quaternion $\quat{\inertiali(\tk)}{\earthi(\tk)}$ with a gyroscope strapdown integration quaternion $\quat{\imui(\tk)}{\inertiali(\tk)}$, and the 9D estimate $\quat{\imui(\tk)}{\earth}$ as the concatenation of a heading correction rotation $\quat{\earthi(\tk)}{\earth}$, represented by the scalar heading offset $\delta_i(\tk)$, with the aforementioned 6D estimate, i.e.,
{\setlength{\mathindent}{0cm}
\begin{align}
\underbrace{\quat{\imui(\tk)}{\earth}}_{\text{\!\!9D estimate\!\!}} =
\overbrace{\underbrace{\quataa{\delta_i(\tk)}{\inlinevec{0,0,1}}}_{\quat{\earthi(\tk)}{\earth}}}^{\substack{\text{magnetometer}\\ \text{correction}}}
\quatmult
\underbrace{\overbrace{\quat{\inertiali(\tk)}{\earthi(\tk)}}^{\substack{\text{\!\!\!\!accelerometer\!\!\!\!}\\ \text{correction}}}
\quatmult
\overbrace{\quat{\imui(\tk)}{\inertiali(\tk)}}^{\substack{\text{strapdown}\\ \text{\!integration\!}}}}_{\text{6D estimate }}.\label{eq:vqf_state}
\end{align}}
The introduced auxiliary coordinate system $\inertiali(\tk)$, with $\inertiali(t_0)=\imui(t_0)$, is an almost-inertial frame that slowly drifts around arbitrary axes due to errors in gyroscope strapdown integration.
See \autoref{fig:vqf_coordinate_systems} for an illustration of the four distinct coordinate systems that are used in \autoref{eq:vqf_state}.

As demonstrated in \cite{seel2017eliminating}, a drawback of many existing methods is that magnetic disturbances can severely impact the inclination estimates.
While previous methods \cite{valenti2015keeping,seel2017eliminating} have already ensured that the magnetometer correction can only influence the heading but not the inclination, the proposed modular state representation makes this property very explicit by representing the heading offset with a scalar variable $\delta_i(\tk)$ and facilitates simultaneous 6D and 9D orientation estimation.

\begin{figure*}[t]
\centering
\includegraphics{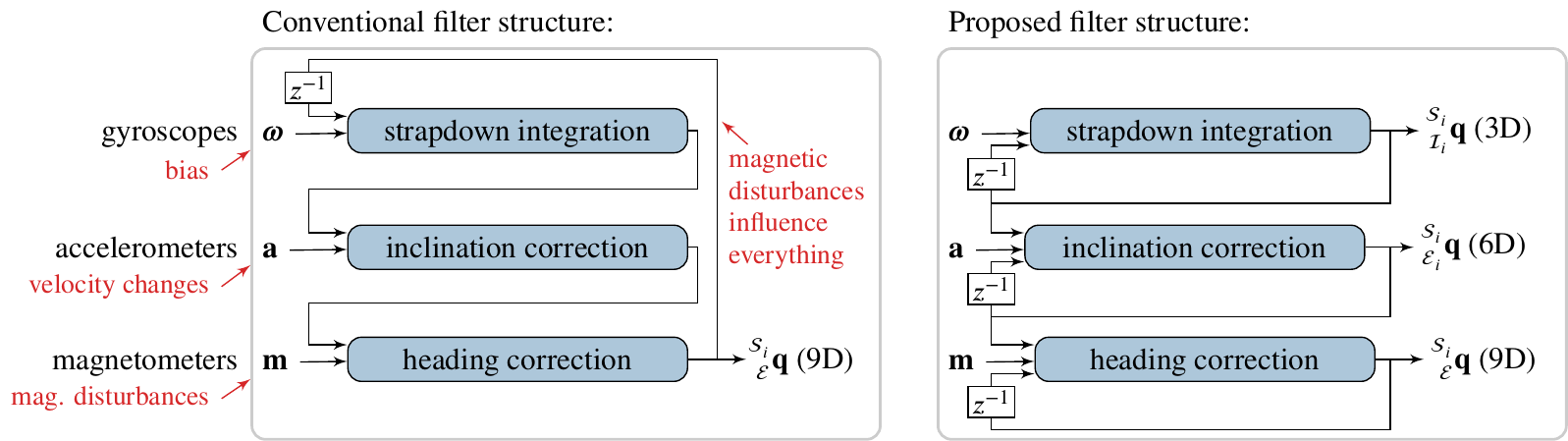}
\caption{Illustration of conventional and proposed filter structures ($z^{-1}$ denotes the unit delay). The proposed filter structure avoids the feedback of the 9D estimate on the strapdown integration. It thereby enables simultaneous 6D and 9D orientation estimation and ensures that the inclination cannot be influenced by magnetic disturbances.}
\label{fig:vqf_filter_structure_comparison}
\end{figure*}

The chosen approach of separating strapdown integration, inclination correction, and heading correction in the state is also represented in the filter structure as shown in \autoref{fig:vqf_filter_structure_comparison}.
Unlike conventional methods, the correction steps are decoupled from the previous steps, i.e., there is no feedback loop from the heading correction to the strapdown integration and, therefore, neither to the inclination correction.

\begin{figure*}[t]
\centering
\includegraphics{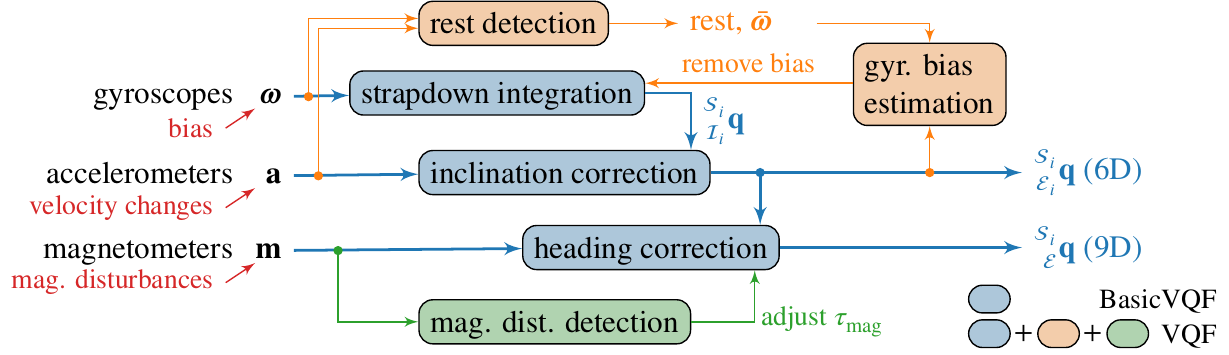}
\caption{Variants of the proposed algorithm. \emph{BasicVQF} consists of strapdown integration, inclination correction, and heading correction. The full version \emph{VQF} additionally includes rest detection, gyroscope bias estimation, and magnetic disturbance rejection (which can be enabled or disabled independently).}
\label{fig:vqf_filter_structure_full}
\end{figure*}

This basic filter structure is extended by an optional gyroscope bias estimation algorithm and an algorithm for magnetic disturbance detection and rejection.
The bias estimation algorithm includes a rest detection and automatically adjusts to whether the \gls{IMU} is at rest or in motion.
The extended filter structure is shown in \autoref{fig:vqf_filter_structure_full}.
Note that it is also possible, and supported by the reference implementation (\autoref{sec:vqf_implementation}), to independently enable or disable rest bias estimation, motion bias estimation, and magnetic disturbance rejection.

In the following, we call the extended algorithm \emph{VQF} (Versatile Quaternion-based Filter) and the basic version \emph{\mbox{BasicVQF}}.
Furthermore, we introduce an acausal implementation called \emph{OfflineVQF} in \autoref{sec:vqf_offline}.

\subsection{Fusion of Gyroscope, Accelerometer, and Magnetometer Measurements}\label{sec:basicvqf}

\begin{figure}[b!]
\centering
\includegraphics{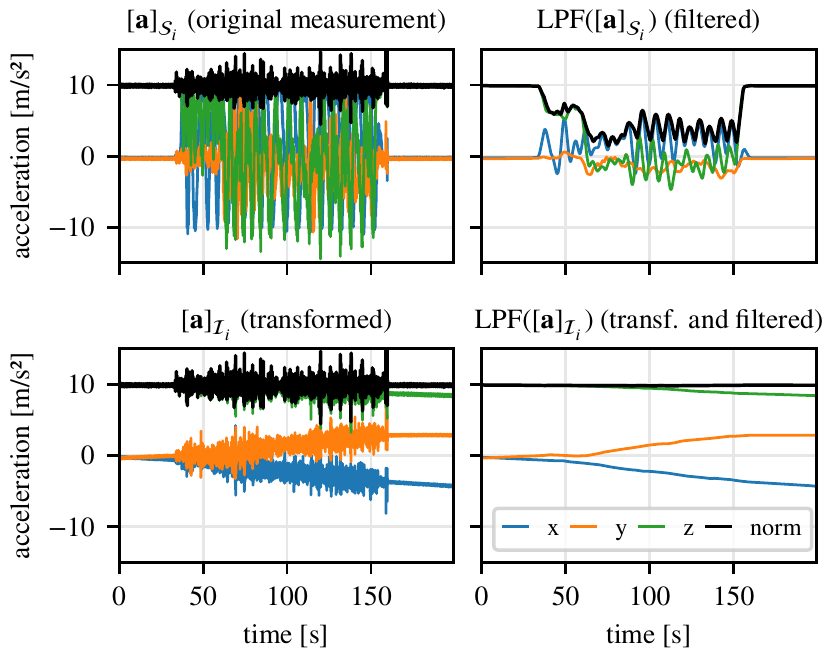}
\caption{Example of unfiltered and low-pass filtered accelerations in the original sensor frame $\imui$ and the almost-inertial frame $\inertiali$. Applying a low-pass filter with a low cutoff frequency to each component in the sensor frame does not give a meaningful output. In the $\inertiali$ frame, low-pass filtering each component of the acceleration effectively averages the measurement, allowing for acceleration and deceleration to cancel out, and the result shows the drift of the $\inertiali$ frame due to errors in gyroscope integration.}
\label{fig:vqf_accplot}
\end{figure}

The basic filter update consists of gyroscope-based prediction, followed by accelerometer correction and, in 9D \gls{IOE}, by magnetometer correction.
The algorithm is given in \sref{Algorithm}{alg:basicvqf}, and details on each step are given in \autoref{app:vqf_basic_step}.
Gyroscope prediction is performed via strapdown integration of the measured angular rate. Errors due to gyroscope bias, noise, and other measurement errors lead to slow drift of the $\inertiali$ frame.

To obtain a vertical reference, we transform the measured accelerations into the almost-inertial frame $\inertiali$ and then apply a second-order Butterworth low-pass filter to each component.
This low-pass filter effectively averages the accelerometer measurements and allows for short-term acceleration and deceleration to cancel out, as illustrated in \autoref{fig:vqf_accplot}.
The inclination of the orientation estimate is then corrected so that the filtered acceleration points in upward direction.
In contrast to conventional \gls{IOE} algorithms, which typically regard each single accelerometer sample as a 3D vector and perform a nonlinear correction step based on the comparison of this vector to a vertical reference vector, the use of a linear low-pass filter in the $\inertiali$ frame more effectively and robustly separates the gravitational acceleration component from the acceleration caused by velocity changes.

If magnetometer measurements are given, a heading offset is derived from the projection of the magnetic field vector into the horizontal plane and tracked via an exponential filter.

\begin{figure*}[htb]
\centering
\begin{minipage}{0.8\textwidth}
\begin{algorithm}[H]
\caption{BasicVQF}\label{alg:basicvqf}
\begin{algorithmic}[1]
    \Procedure{InitializeFilter}{}
        \State $\quat{\imui}{\inertiali} \gets \inlinevec{1,0,0,0}$ \Comment{Gyroscope strapdown integration quaternion}
        \State $\quat{\inertiali}{\earthi} \gets \inlinevec{1,0,0,0}$ \Comment{Accelerometer correction quaternion}
        \State $\delta_i \gets 0$ \Comment{Magnetometer correction angle}
        \State initialize low-pass filter state
    \EndProcedure
    \Procedure{FilterUpdate}{$\gyr{}=\gyr{}(\tk),\nvec{a}{}=\nvec{a}{}(\tk),\nvec{m}{}=\nvec{m}{}(\tk), f_\mathrm{c,acc}, k_\mathrm{mag}, \Ts$}
        \State $\quat{\imui}{\inertiali}\gets \quat{\imui}{\inertiali} \quatmult \quataa{\Ts \norm{\gyr{}{}}}{\gyr{}{}}$\Comment{Perform gyroscope strapdown integration}
            \State $\nvec[\inertiali]{a}{} \gets \quat{\imui}{\inertiali}\quatmult\nvec{a}{}\quatmult\quat{\imui}{\inertiali}\quatinv$\Comment{Transform acceleration to $\inertiali$ frame}
            \State $\nvec[\inertiali]{a}{LP} \gets \operatorname{lpfStep}(\nvec[\inertiali]{a}{}, f_\mathrm{c}=f_\mathrm{c,acc})$\Comment{Apply low-pass filter}
            \State $\nvec[\earthi]{a}{LP} \gets \quat{\inertiali}{\earthi}\quatmult\nvec[\inertiali]{a}{LP}\quatmult\quat{\inertiali}{\earthi}\quatinv$\Comment{Transform to \earthi{} frame}
            \State $\inlinevec{a_x,a_y,a_z} \gets \normalized{\nvec[\earthi]{a}{LP}}$\Comment{Normalize}
            \State $q_w \gets \sqrt{\frac{a_z + 1}{2}}$
            \State $\quat{\inertiali}{\earthi} \gets \inlinevec{q_w,\frac{a_y}{2 q_w},\frac{-a_x}{2 q_w},0} \quatmult \quat{\inertiali}{\earthi}$\Comment{Update correction quaternion}
        \State $\quat{\imui}{\earthi} \gets \quat{\inertiali}{\earthi}\quatmult\quat{\imui}{\inertiali}$\Comment{Calculate 6D orientation estimate}
        \If{\nvec{m}{} is given}%
            \State $\inlinevec{m_x, m_y, m_z} \gets \quat{\imui}{\earthi}\quatmult\nvec{m}{}\quatmult\quat{\imui}{\earthi}\quatinv$\Comment{Transform mag. sample to $\earthi$ frame}
            \State $\delta_\mathrm{mag} \gets \atantwo(m_x, m_y)$\Comment{Calculate heading offset from mag. sample}
            \State $\delta_i \gets \delta_i + k_\mathrm{mag} \operatorname{wrapToPi}(\delta_\mathrm{mag} - \delta_i)$\Comment{Update correction angle}
        \EndIf
        \State $\quat{\imui}{\earth} \gets \inlinevec{\cos\frac{\delta_i}{2},0,0,\sin\frac{\delta_i}{2}}\quatmult\quat{\imui}{\earthi}$\Comment{Calculate 9D orientation estimate}
        \State \textbf{return} $\quat{\imui}{\earthi}$, $\quat{\imui}{\earth}$\Comment{Provide 6D and 9D orientation estimate}
    \EndProcedure
\end{algorithmic}
{\footnotesize
lpfStep: update step of second-order Butterworth low-pass filter with cutoff frequency $f_\mathrm{c}$\\
wrapToPi: bring angle into the interval $[-\pi,\pi]$ by adding integer multiples of $2\pi$
}
\end{algorithm}
\end{minipage}
\end{figure*}

As it is common in \gls{IOE} algorithms, the behavior can be influenced by fusion weights that balance between rejecting gyroscope drift and rejecting disturbances in the accelerometer and magnetometer measurements.
In \sref{Algorithm}{alg:basicvqf}, those parameters are the cut-off frequency $f_\mathrm{c,acc}$ and the magnetometer correction gain $k_\mathrm{mag}$.
Like for many existing methods, the meaning of the values assigned to those parameters is hard to interpret, depends on the sampling time (for $k_\mathrm{mag}$), and does not allow for a comparison between the trust assigned to the accelerometer and the trust assigned to the magnetometer.
To provide a more intuitive parametrization, we replace those parameters with time constants $\tauAcc$ and $\tauMag$ that can be changed by the user to influence the algorithm behavior.
A small time constant leads to fast correction and indicates high trust in the accelerometer or magnetometer measurements, while large time constants indicate trust in the gyroscope measurements.

Those time constants map to the internal values as follows (see \autoref{app:vqf_basic_step} and \autoref{fig:vqf_step_response_tau_plot} for more information).
For the magnetometer correction first-order exponential filter, we use the time constant that is commonly used to characterize first-order systems and corresponds to the time needed for the step response to reach $1-e^{-1} \approx \SI{63.2}{\percent}$ of its final value.
To ensure similar behavior for the second-order Butterworth filter of the accelerometer correction, we use a time constant that corresponds to the undampened part of the step response.
This leads to the mapping
\begin{equation}
f_\mathrm{c,acc}=\frac{\sqrt{2}}{2\pi\tauAcc}, \quad
k_\mathrm{mag} = 1 - \exp\left(-\frac{\Ts}{\tauMag}\right),
\end{equation}
which allows us to derive the internal parameters $f_\mathrm{c,acc}$ and $k_\mathrm{mag}$ from the user-specified time constants $\tauAcc$ and $\tauMag$.
The same parametrization via time constants is also used for the other first-order and second-order filters introduced in the following subsections.
Note that we will later determine default values for $\tauAcc$ and $\tauMag$ that yield excellent out-of-the-box accuracy for a large range of application scenarios, and manual tuning by adjusting these time constants is therefore only required in rare edge cases.

\subsection{Gyroscope Bias Estimation}

\begin{figure*}[htb]
\centering
  \includegraphics{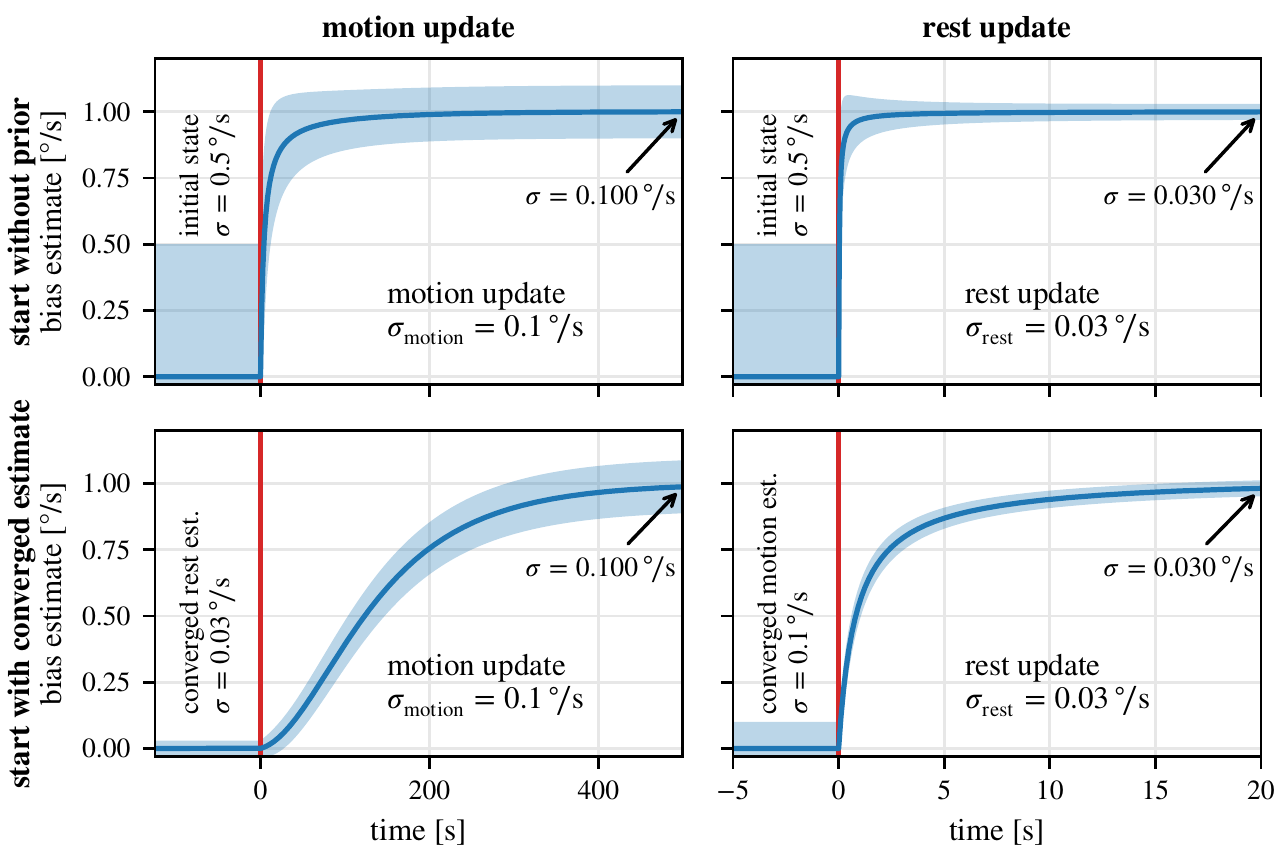}
  \caption{Step response for the Kalman filter with the proposed parametrization. The blue bands show the standard deviation $\sigma$ of the estimate. At $t<0$, the Kalman filter either starts in the initial state ($\sigma=\SI{0.5}{\dps}$) or in a converged state of either the motion or rest update with an estimate of \SI{0}{\dps}. At $t=0$, the measurement changes to \SI{1}{\dps}. Directly after initialization, the filter converges much faster than in cases where a (contradictory) previous estimate was already obtained. In general, the motion update converges much slower than the rest update.}
  \label{fig:vqf_KF_convergence_plot}
\end{figure*}

To ensure high accuracy in the presence of gyroscope bias, we extend the BasicVQF algorithm from \autoref{sec:basicvqf} by a method to estimate and compensate such bias.
In existing \gls{IOE} algorithms, this is commonly realized by integral action \cite{seel2017eliminating,mahony2008nonlinear,madgwick2010efficient}.
However, this approach requires feedback of both accelerometer and magnetometer correction \cite{madgwick2010efficient}, making the gyroscope bias estimate susceptible to magnetic disturbances.
To prevent this, the proposed method for gyroscope bias estimation avoids using any information from the magnetometer correction.
Instead, the bias is estimated solely from the disagreement between strapdown integration and accelerometer measurements during motion.

Moreover, note that, in many application scenarios, the \gls{IMU} will occasionally be at rest for several seconds, and those phases can be detected reliably.
This is leveraged by the proposed bias estimation algorithm, which determines the bias directly from low-pass-filtered gyroscope measurements whenever it detects a period of rest.

As further detailed in \autoref{app:vqf_bias_est} and \autoref{alg:biasest}, the bias estimation is realized via a single Kalman filter that performs different updates based on whether the \gls{IMU} is currently at rest or in motion.
This Kalman filter is parametrized in a way that is independent of the sampling rate and that gives much larger trust to the rest-based updates than to the motion-based updates.
In each sampling step, the current Kalman filter estimate of the bias is subtracted from the gyroscope measurement before strapdown integration.

\autoref{fig:vqf_KF_convergence_plot} shows how the Kalman filter behaves with the proposed parametrization.
The relation between the uncertainty of the measurement (large during motion, small during rest) and the uncertainty of the current estimate determines how fast the bias estimation converges.
After initialization without prior knowledge, the estimation uncertainty is large, which leads to fast convergence.
From a converged estimate during motion (i.e., with medium uncertainty), a contradicting but much more reliable observation during rest is adopted within several seconds.
From a converged rest estimate, a contradicting observation during motion is adopted much slower due to the larger uncertainty of the measurement.

\newpage

\subsection{Magnetic Disturbance Rejection}\label{sec:vqf_mag_dist_rejection}

\begin{table*}[t!]
\small
\caption{Inertial orientation estimation algorithms used in the evaluation}\label{tab:vqf_algorithm_overview}
\begin{tabular}{llllll}
\toprule
\multicolumn{2}{l}{\textbf{Algorithm}} & \textbf{6D} & \textbf{9D} & \textbf{Bias} & \textbf{Source}\\\midrule
\textbf{VQF}
 & proposed method
 & \textcolor{C2dark}{\cmark}
 & \textcolor{C2dark}{\cmark}
 & \textcolor{C2dark}{\cmark}
 & \scriptsize \url{https://github.com/dlaidig/vqf}
\\[0.3em]
MAH
 & \citet{mahony2008nonlinear}
 & \textcolor{C2dark}{\cmark}
 & \textcolor{C2dark}{\cmark}
 & \textcolor{C2dark}{\cmark}
 & \scriptsize \url{https://x-io.co.uk/open-source-imu-and-ahrs-algorithms/}
\\[0.3em]
MAD
 & \citet{madgwick2010efficient}
 & \textcolor{C2dark}{\cmark}
 & \textcolor{C2dark}{\cmark}
 & \textcolor{C3!30!white}{\xmark}
 & \scriptsize \url{https://x-io.co.uk/open-source-imu-and-ahrs-algorithms/}
\\[0.3em]
VAC
 & \citet{valenti2015keeping}
 & \textcolor{C2dark}{\cmark}
 & \textcolor{C2dark}{\cmark}
 & \textcolor{C2dark}{\cmark}
 & \scriptsize \url{https://wiki.ros.org/imu_complementary_filter}
\\[0.3em]
FKF
 & \citet{guo2017novel}
 & \textcolor{C3!30!white}{\xmark}
 & \textcolor{C2dark}{\cmark}
 & \textcolor{C3!30!white}{\xmark}
 & \scriptsize \url{https://github.com/zarathustr/FKF}
\\[0.3em]
SEL
 & \citet{seel2017eliminating}
 & \textcolor{C2dark}{\cmark}
 & \textcolor{C2dark}{\cmark}
 & \textcolor{C2dark}{\cmark}
 & \scriptsize \url{https://github.com/dlaidig/qmt}, qmt.oriEstIMU
\\[0.3em]
MKF
 & Matlab
 & \textcolor{C2dark}{\cmark}
 & \textcolor{C2dark}{\cmark}
 & \textcolor{C2dark}{\cmark}
 & \scriptsize Matlab R2021b (The MathWorks Inc., Natick, MA, USA), imufilter/ahrsfilter
\\[0.3em]
KOK
 & \citet{kok2019fast}
 & \textcolor{C3!30!white}{\xmark}
 & \textcolor{C2dark}{\cmark}
 & \textcolor{C2dark}{\cmark}
 & \scriptsize \url{https://github.com/manonkok/fastRobustOriEst}
\\[0.3em]
RIANN
 & \citet{weber2021riann}
 & \textcolor{C2dark}{\cmark}
 & \textcolor{C3!30!white}{\xmark}
 & \textcolor{C3!30!white}{\xmark}
 & \scriptsize \url{https://github.com/daniel-om-weber/riann}
\\\bottomrule
\end{tabular}
\end{table*}

We extend the proposed \gls{IOE} algorithm with a set of methods that enable adaptive filtering of the magnetometer measurements with the aim of reducing the influence of temporary magnetic disturbances.
The employed strategy is composed of three parts: magnetic disturbance \emph{detection}, magnetic disturbance \emph{rejection}, and new magnetic field \emph{acceptance}.
All three parts are briefly explained below, and the complete algorithm is described in detail in \autoref{alg:magdist} in \autoref{app:vqf_mag_dist_rejection}.

The magnetic disturbance \emph{detection} uses a user-defined or automatically determined reference for the norm and dip angle of the local magnetic field.
The magnetometer measurements are considered to be undisturbed only if they have been close to the reference for at least $\SI{0.5}{\second}$ -- and disturbed otherwise.
Whenever the magnetic field is considered undisturbed, the reference values are slowly updated to track slow changes in the norm and dip angle.

The magnetic disturbance \emph{rejection} adjusts the speed of the first-order filter used for heading correction if magnetic disturbances are detected.
For disturbances of up to \SI{60}{\second}, the magnetometer update is fully disabled.
For longer periods that are considered to be disturbed, updates are performed but at a lower speed.

Finally, the new magnetic field \emph{acceptance} is used to deal with sudden changes in the environment, e.g., after changing the terrain from outdoor to indoor or moving to a different indoor room with a different local magnetic field.
Whenever the magnetic field is considered to be disturbed but seems homogeneous (for a sufficiently long time during which the \gls{IMU} was not stationary), the norm and dip angle of the current magnetic field is used as the new reference.

\subsection{Acausal Filtering for Offline Data Processing}\label{sec:vqf_offline}

In application scenarios in which the complete time series of recorded data is available (offline data processing), we can employ acausal signal processing methods to further improve accuracy.
As commonly done in signal processing for zero-phase filtering \cite{gustafsson1996determining}, we first run the filtering steps forward and then again backward in time.
This allows us to leverage the existing real-time implementation to create the offline variant detailed in \autoref{alg:vqf_offline} in \autoref{app:vqf_offline}.

\subsection{Open-Source Implementation}\label{sec:vqf_implementation}

Implementations of the proposed orientation estimation algorithm are available at \url{https://github.com/dlaidig/vqf} under the MIT license.
Native implementations are provided in C++, Python, and Matlab.
Furthermore, the fast C++ implementation can easily be used from Python code.
The Python package is available at \url{https://pypi.org/project/vqf/} and can be installed via pip, and documentation is available at \url{https://vqf.readthedocs.io/}.

\section{Evaluation}

In this section, we evaluate the performance of the proposed \gls{IOE} algorithm on six publicly available datasets and compare the results obtained with the proposed method to results obtained with eight state-of-the-art \gls{IOE} algorithms.

\begin{figure*}[t]
\centering
\includegraphics{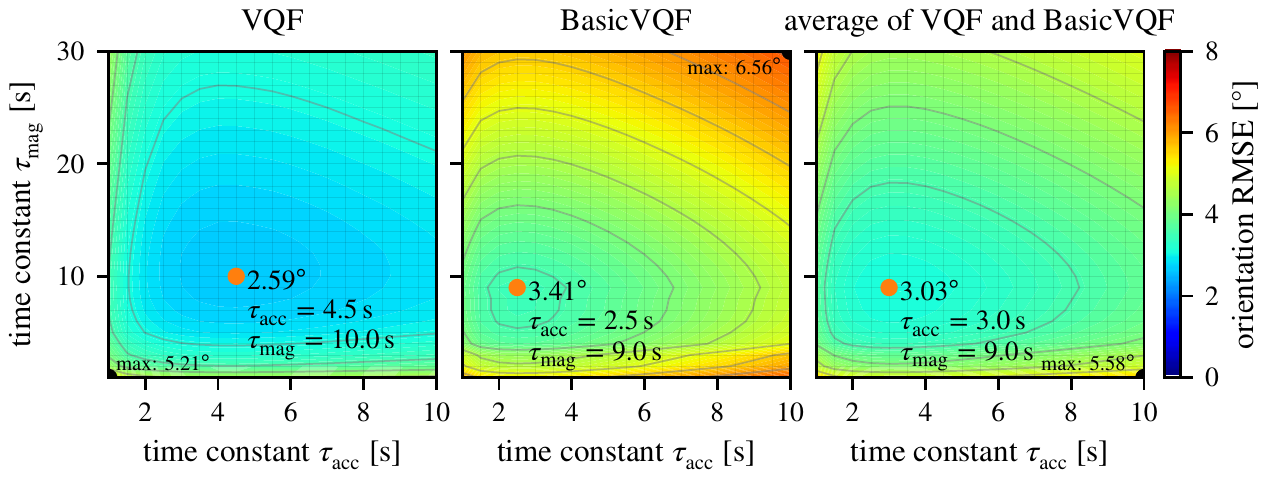}
\caption{RMSE (weighted average over all datasets) achieved with the proposed VQF algorithm and with the reduced BasicVQF variant, for different values of the tuning parameters. The default algorithm parameters are chosen such that the mean of both errors is minimized, i.e., $\tauAcc=\SI{3}{\second}$ and $\tauMag=\SI{9}{\second}$.}
\label{fig:vqf_tagpx_contour_plot}
\end{figure*}

\begin{figure*}[t]
\centering
\includegraphics{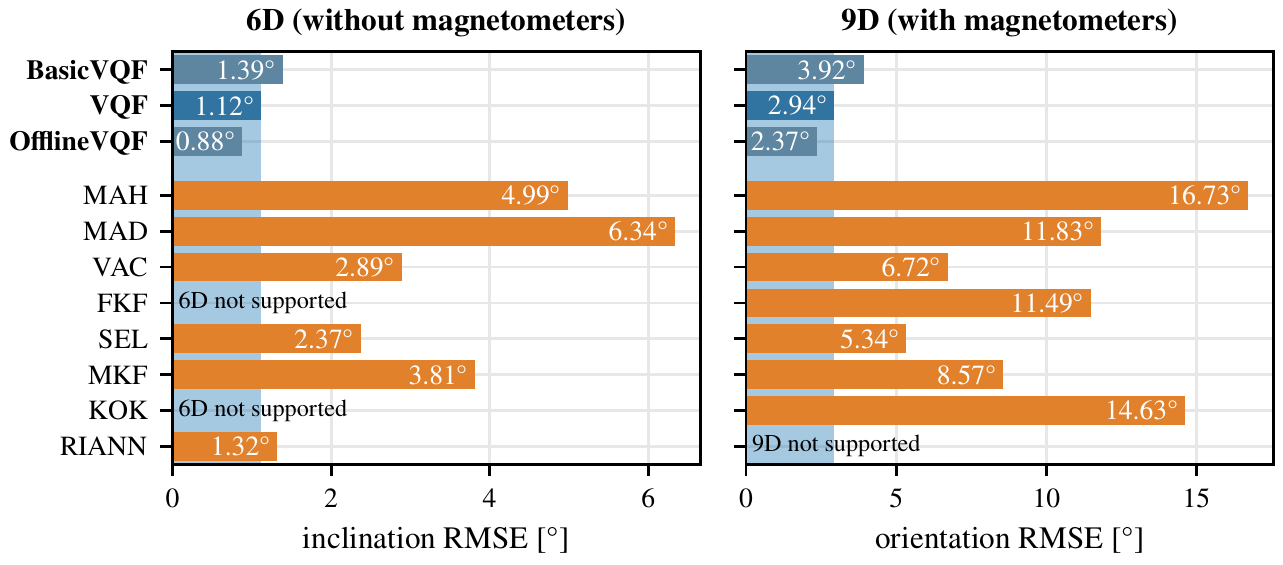}
\caption{RMSE (weighted average over all datasets) for the proposed VQF and all state-of-the-art algorithms. VQF outperforms all eight literature methods, and only RIANN provides similar 6D performance. Even the errors obtained with the simple BasicVQF variant are clearly lower than for the other seven algorithms. When real-time capability is not required, using the offline variant is advised to further increase the accuracy of the orientation estimates.}
\label{fig:vqf_simple_method_comparison_plot}
\end{figure*}

\begin{figure*}[t!]
\includegraphics{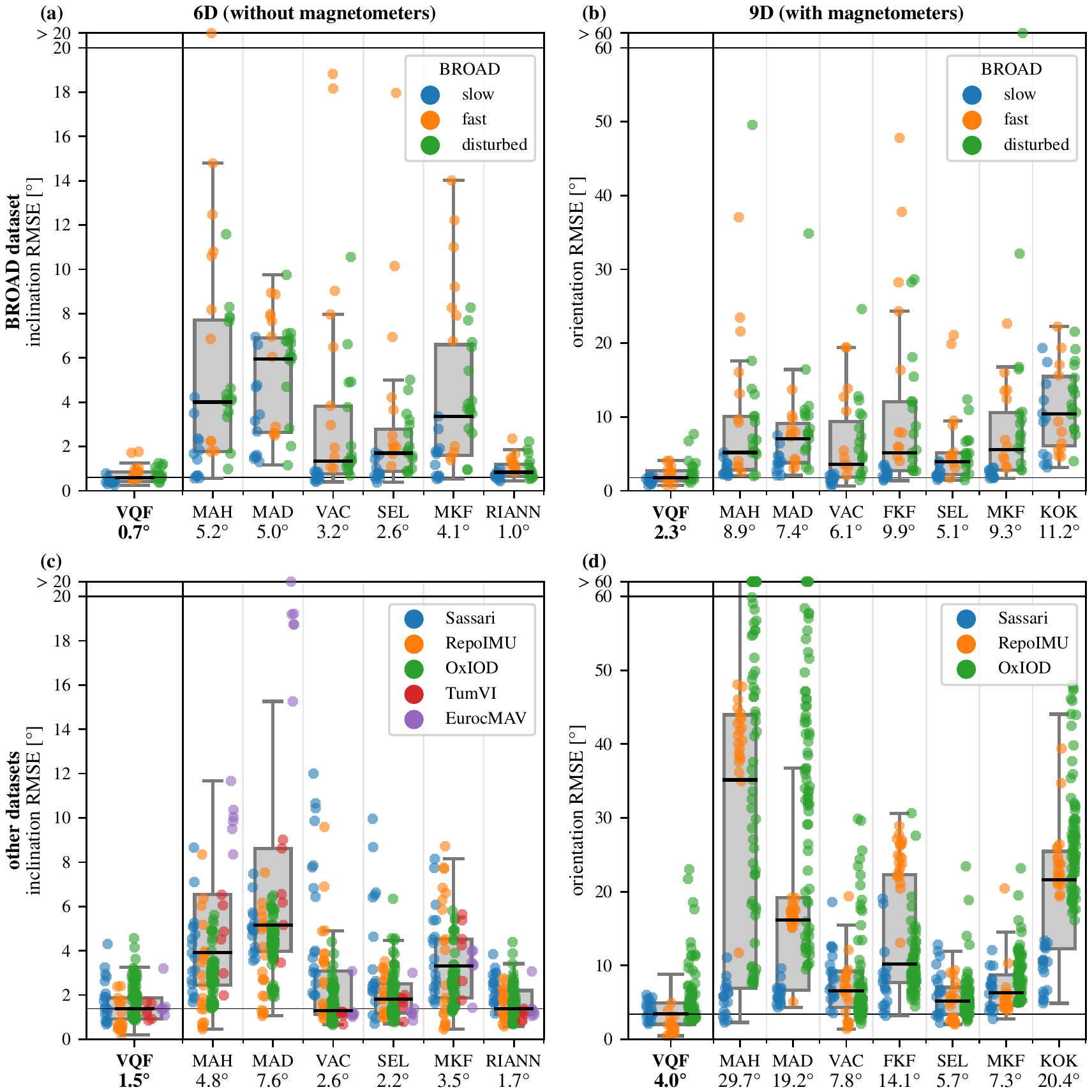}
{\phantomsubcaption\label{fig:vqf_method_comparison_plotA}}
{\phantomsubcaption\label{fig:vqf_method_comparison_plotB}}
{\phantomsubcaption\label{fig:vqf_method_comparison_plotC}}
{\phantomsubcaption\label{fig:vqf_method_comparison_plotD}}
\caption{Orientation estimation errors for all evaluated algorithms and for all trials of the \textbf{(a,b)} BROAD dataset and  \textbf{(c,d)} the five other datasets and for \textbf{(a,c)} 6D and \textbf{(b,d)} 9D sensor fusion. The numbers below the algorithm names indicate the RMSE averaged over all trials. For \textbf{(b,d)}, the boxplots and average values are weighted to give each dataset the same weight regardless of the number of trials. The proposed VQF algorithm consistently provides the best performance.}
\label{fig:vqf_method_comparison_plot}
\end{figure*}

\begin{figure*}[t]
\includegraphics{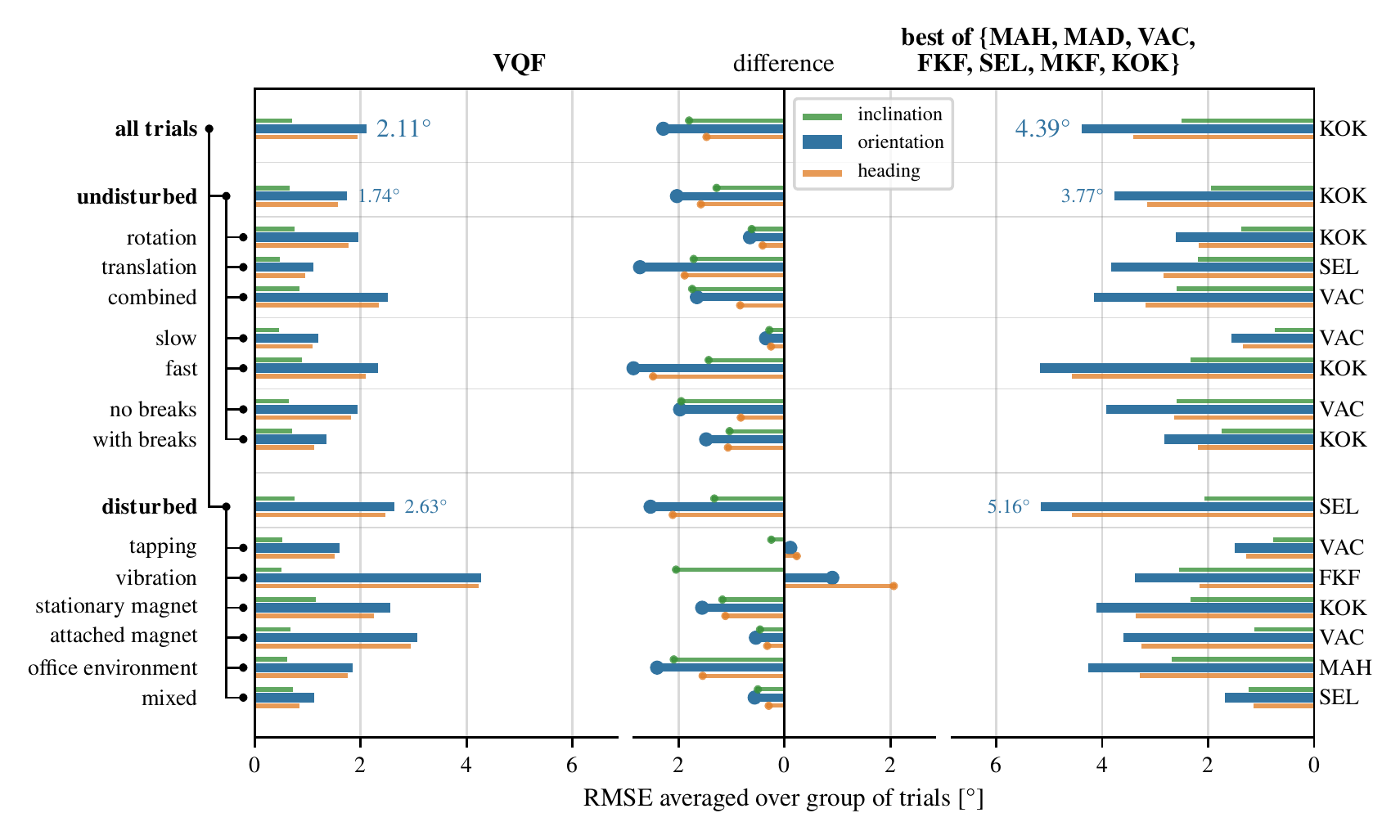}
\caption{Averaged RMSE errors for various groups of trials of the BROAD dataset. The proposed algorithm VQF is compared with the best of the seven other evaluated algorithms, i.e., the algorithm that provides the lowest orientation error for the respective group of trials. The lines originating from the center highlight the difference between the errors. For all groups except for the \emph{tapping} and \emph{vibration} trials, the proposed algorithm outperforms even the best-performing literature method.}
\label{fig:vqf_group_comparison_plot}
\end{figure*}

\subsection{Algorithms and Datasets}

\autoref{tab:vqf_algorithm_overview} lists the algorithms used for the evaluation.
This table also shows whether the algorithm can be used for 6D and/or 9D orientation estimation, whether it supports gyroscope bias estimation, and where the implementation is available.

To evaluate the accuracy of \gls{IOE}, we consider publicly available datasets consisting of \gls{IMU} measurements and a ground truth orientation obtained from marker-based \acrfull{OMC}.
In \cite{laidig2021broad}, we presented the \acrfull{BROAD} and briefly reviewed other existing datasets that contain trial data that is suitable for \gls{IOE} accuracy evaluation.
In the present evaluation, we use all of these datasets, i.e.,
\begin{itemize}
\item BROAD \cite{laidig2021broad}: 39 trials (23 undisturbed trials with different motion types and speeds and 16 trials with various deliberate disturbances)
\item Sassari \cite{caruso2021orientation}: 18 trials (3 speeds, 3 \gls{IMU} models, and 2 \glspl{IMU} of each model)
\item RepoIMU \cite{szczesna2016reference}: 21 trials (\emph{T-Stick} only; test 5, test 6 trial 1, test 10 excluded due to artifacts, as explained in \cite{laidig2021broad})
\item OxIOD \cite{chen2018oxiod}: 71 trials (only \emph{handbag}, \emph{handheld}, \emph{pocket}, \emph{running}, \emph{slow walking}, \emph{trolley} trials)
\item TUM VI \cite{schubert2018tum}: 6 trials (\emph{room} only; no magnetometer data)
\item EuRoC MAV \cite{burri2016euroc}: 6 trials (\emph{Vicon room} only; no magnetometer data).
\end{itemize}
Combined, the collection of evaluation data consists of 161 trials with a total duration of \SI{12.9}{\hour}.
The data includes motions of handheld \glspl{IMU} (various combinations of fast and slow rotations and translations), walking and running, as well as flight data from a micro aerial vehicle, and contains data from eight different \gls{IMU} models, recorded at sampling rates ranging from \SI{100}{\hertz} to \SI{286}{\hertz}.
For more information about the datasets, refer to the respective publications and the summary provided in \cite{laidig2021broad}.
This large collection of experimental data allows us to evaluate the robustness of the proposed method for different motion characteristics, different sensor hardware, and different sampling rates.

\subsection{Algorithm Parametrization}\label{sec:parametrization}

Before assessing the \gls{IOE} accuracy, we need to determine suitable tuning parameters for each algorithm.
In \cite{laidig2021broad}, we introduced a metric to assess the performance of an \gls{IOE} algorithm, the \gls{TAGP}.
This metric is defined as the smallest possible \gls{RMSE}, averaged over all 39 trials of the \gls{BROAD} dataset, that can be obtained with a common algorithm parameterization for all trials.
The associated parameterization can then be expected to provide good results for a wide variety of motions and disturbance scenarios.

However, one limitation of the \gls{BROAD} dataset is that all trials are performed with the same \gls{IMU} model.
To find parameters that are not only robust against movement speed, type of motion, and various disturbances, but also work well for different sensor characteristics, we define an extended \gls{TAGP} metric, the \tagpx.

Similar to the \gls{TAGP}, the \tagpx{} is the smallest possible \acrshort{RMSE}, averaged over the aforementioned trials for each of the six datasets.
The errors are first averaged by dataset, where the dedicated benchmark dataset \gls{BROAD} is given a five times larger weight than all other datasets, which are weighted equally.
For trials without magnetometer data (and for RIANN, which does not support magnetometers), the inclination error, as defined in \cite{laidig2021broad}, is used instead of the orientation error.

\autoref{fig:vqf_tagpx_contour_plot} shows how the weighted error defined above depends on the tuning parameters for the default and the basic variant of the proposed VQF algorithm.
The \tagpx{} is the minimum value of this error, i.e., \ang{2.59} for VQF and \ang{3.41} for BasicVQF, which shows that gyroscope bias estimation and magnetic disturbance rejection lead to improved accuracy.
However, the optimal values for the time constants \tauAcc{} and \tauMag{} are similar for both variants.
To avoid specifying different default parameters for VQF and BasicVQF, we simply use the average of the errors obtained with both variants to determine the default values $\tauAcc=\SI{3}{\second}$ and $\tauAcc=\SI{9}{\second}$.

To provide a fair comparison between the proposed and the state-of-the-art methods, we also optimize the parameters for all other algorithms according to the \tagpx{}, i.e., we find the parameters that allow each algorithm to provide the best possible performance across all datasets.
Details on the employed search strategy as well as the resulting parameters are given in \autoref{app:vqf_evaluation_parameter_tuning}.
In the following, we always use those optimal parameters to evaluate and compare algorithm performance.

\begin{figure*}[t]
\centering
\includegraphics{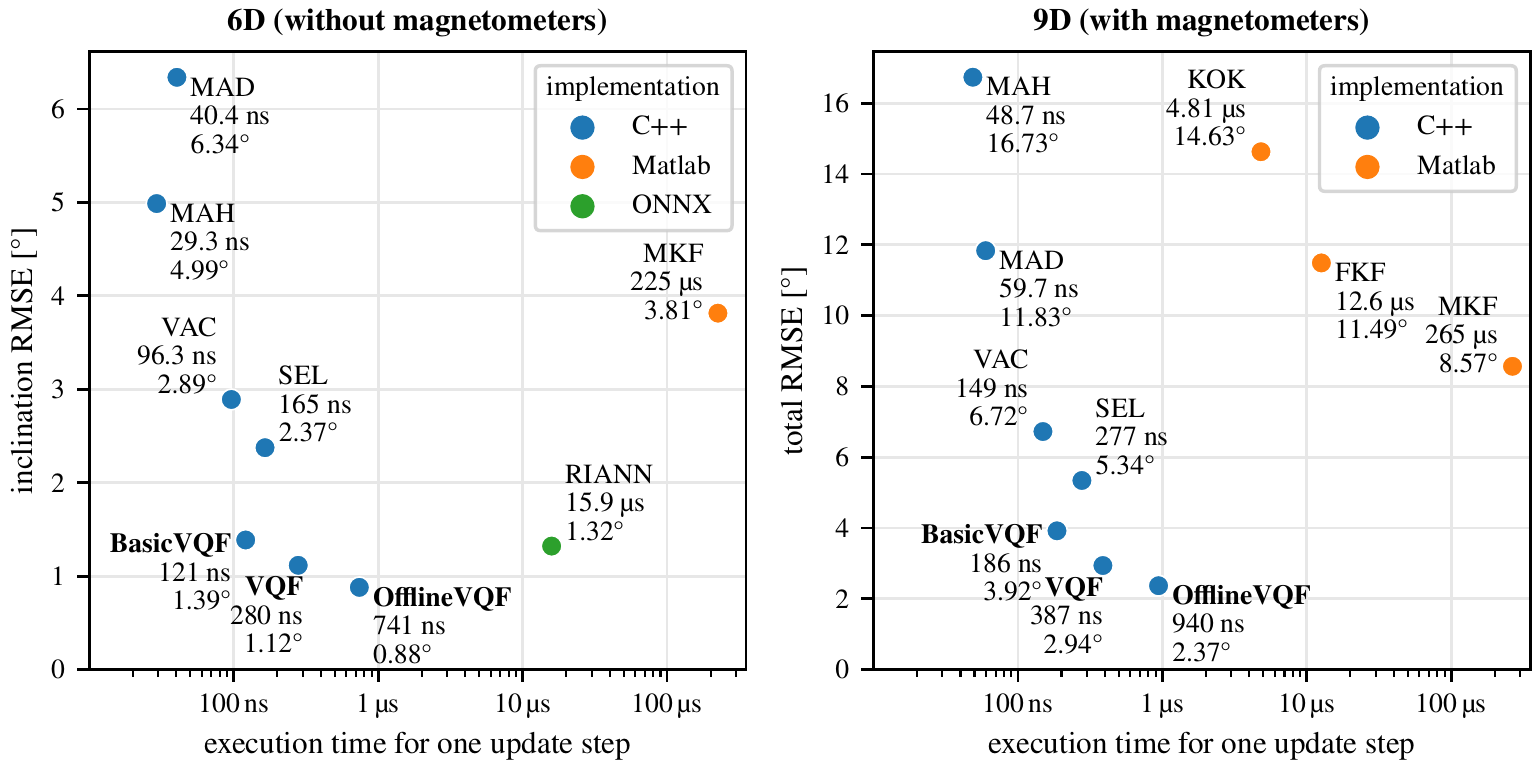}
\caption{Execution time for one update step on an AMD Ryzen 5 3600 CPU vs. orientation estimation RMSE (weighted average over all datasets) for the proposed VQF and all state-of-the-art algorithms. Accuracies vary largely, and execution times mainly depend on the programming language used for implementing the algorithm. The execution time of VQF is in the same order of magnitude as for the other algorithms implemented in C++, while the errors are clearly smaller.}
\label{fig:vqf_execution_time_vs_error_plot}
\end{figure*}

\subsection{Orientation Estimation Accuracy}

To assess the performance of all algorithms, we apply each to the data of all trials of all datasets.
In the case of 6D (magnetometer-free) sensor fusion, we determine the inclination error \cite{laidig2021broad}, and for 9D sensor fusion the orientation error, in each case between the \gls{IMU}-based estimate and the \gls{OMC} ground truth.
We then calculate the \gls{RMSE} while only considering the motion phases.
Since the TUM VI and EuRoC MAV datasets do not include magnetometer data, they are only considered for the 6D results.
We average the errors by dataset analogously to the definition of the \tagpx{} in \autoref{sec:parametrization}.
The resulting values are presented for all algorithms in \autoref{fig:vqf_simple_method_comparison_plot}.

One main observation from this figure is that the proposed method VQF consistently provides considerably lower errors than the existing orientation estimation algorithms.
For 9D \gls{IOE}, there is a 1.8-fold to 5-fold increase in accuracy, while for 6D \gls{IOE}, the increase is \SI{17}{\percent} for RIANN and between 2.1-fold and 5-fold for the other methods.
It should be noted that the only algorithm that achieves similar inclination errors, the neural network RIANN, cannot perform 9D sensor fusion and was trained on some of the datasets that are used in this evaluation (see \cite{weber2021riann} for details).

\autoref{fig:vqf_simple_method_comparison_plot} also allows us to compare the variants of the proposed method, cf.\ \autoref{fig:vqf_filter_structure_full}.
Unsurprisingly, the errors obtained with the BasicVQF variant (no bias estimation and no magnetic disturbance rejection) are slightly larger.
However, with the aforementioned exception of RIANN, even the 6D and 9D errors of BasicVQF are still clearly lower than the corresponding errors obtained with any of the existing algorithms.
The low 6D errors of BasicVQF can directly be attributed to the novel approach for inclination correction, while the comparatively small difference between 6D and 9D errors can be attributed to the decoupled filter structure and modular state representation.
Compared to the real-time-capable implementation, the offline variant OfflineVQF is able to further increase the estimation accuracy by another 20 percent.
Therefore, employing this variant is advisable when analyzing recorded data.

While \autoref{fig:vqf_simple_method_comparison_plot} shows a clear improvement in accuracy with respect to the state of the art when looking at errors averaged over a large number of trials, a comprehensive comparison should also include a closer look at individual trials.
\autoref{fig:vqf_method_comparison_plot} differentiates the errors by dataset and shows a marker for each single trial.
Comparing medians, the interquartile ranges, the lengths of the whiskers, or the distributions of outliers yields the same conclusion: The proposed method not only performs better on average but consistently and robustly provides lower errors than the existing methods.

It is noticeable that unusually large errors of $\sim\ang{23}$ are observed for some trials of the OxIOD dataset.
As those large errors are observed across all methods, the most likely cause of those errors are irregularities in the measurement or ground truth data that have previously been noted in \cite{laidig2021broad}.

Beyond the level of comparing performance on different datasets, the \gls{BROAD} dataset facilitates the investigation of algorithm performance for seven different motion characteristics and six different disturbance characteristics.
\autoref{fig:vqf_group_comparison_plot} compares the average RMSE across different groups of trials and compares the proposed VQF method in each case with the best of the other 9D-capable algorithms, i.e., the algorithm that achieves the lowest errors for the respective group of trials.
Except for the \emph{tapping} and \emph{vibration} groups, VQF always achieves lower errors than even the best of the other algorithms.
For the \emph{vibration} group, it is worth noticing that, while the orientation error is slightly larger than the error obtained with FKF, the inclination errors obtained with VQF are clearly lower.

In summary, compared with eight other \gls{IOE} algorithms and using a collection of six publicly available datasets that cover a wide range of motions, speeds, disturbances, and different sensor hardware, the proposed method VQF consistently provides the best \gls{IOE} accuracy, both for 6D and 9D orientation estimation.

\subsection{Algorithm Execution Time}

In addition to accuracy, the execution time of an \gls{IOE} algorithm is often relevant, especially in real-time applications or when the algorithm is running on low-powered microcontrollers directly on the \gls{IMU}.
To compare the execution times, we repeatedly process the entire BROAD dataset with all algorithms on an AMD Ryzen 5 3600 CPU while measuring the execution time.
\autoref{fig:vqf_execution_time_vs_error_plot} shows the average execution time for one update step in combination with the orientation estimation error for the respective algorithm.
The results show that execution time mostly depends on the programming language used for the implementation and that the algorithms written in C++ are considerably faster than the algorithms written in Matlab or using the ONNX machine learning runtime.
While the VQF algorithms achieve clearly higher accuracy, the execution times are in the same order of magnitude as for the existing state-of-the-art methods with a C++ implementation.
VQF is fast enough for use on microcontrollers, which we verified by integrating it into an \gls{IMU} firmware running on a Cortex M4 at a comparatively high sampling rate of \SI{1600}{\hertz}.

\begin{figure*}[htb]
\includegraphics{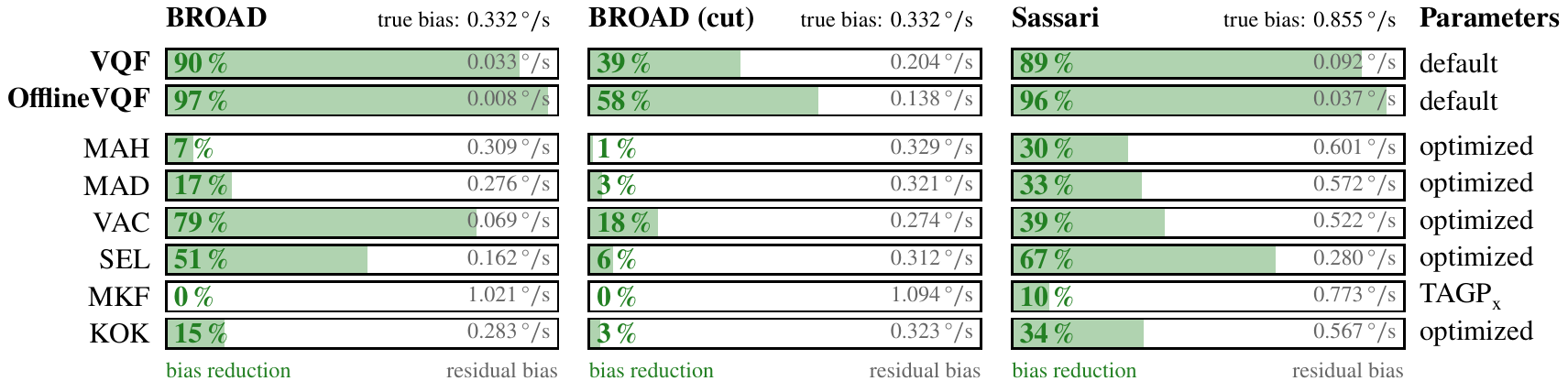}
\caption{Bias estimation results for different IOE algorithms. Green bars and percentages indicate the reduction of the bias norm with respect to the true bias (black) contained in the measurement data.
For all datasets, VQF with default parameters surpasses the best possible performance of the existing algorithms. Using the OfflineVQF variant further improves accuracy.}
\label{fig:vqf_bias_estimation_plot}
\end{figure*}

\begin{figure*}[t]
\includegraphics{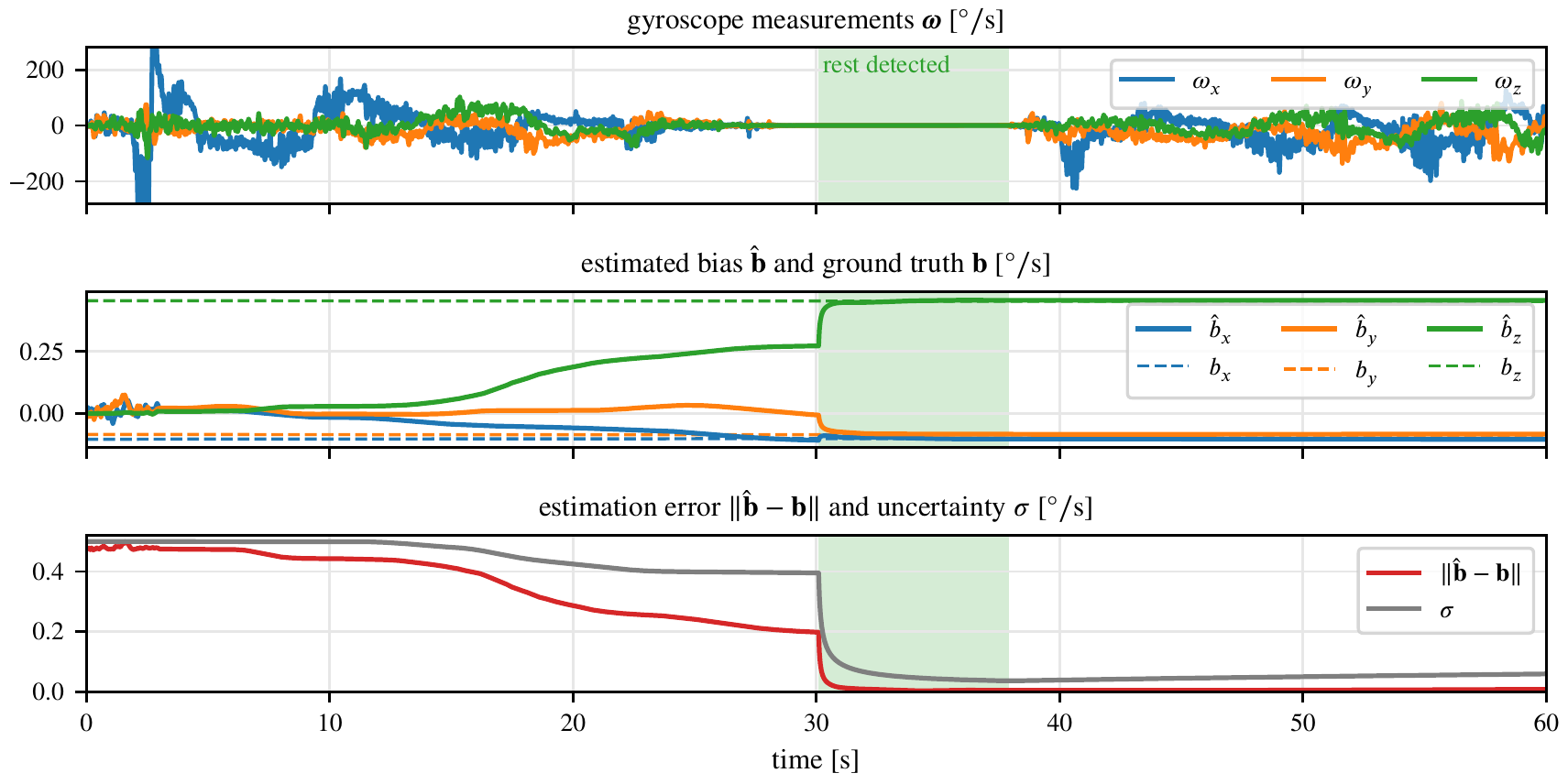}
\caption{Behavior of the bias estimation method for the first 60 seconds of trial 4 of the BROAD dataset (undisturbed slow rotation with breaks, initial rest cut). During motion, the estimated bias slowly converges toward the true value. Once rest is automatically detected by \autoref{alg:biasest}, the convergence speed increases, and the error and the estimation uncertainty suddenly drop.}
\label{fig:vqf_bias_example_plot}
\end{figure*}

\subsection{Gyroscope Bias Estimation}

We now investigate the performance of the gyroscope bias estimation method.
Besides the proposed algorithm, we also evaluate the performance of the five other algorithms that are able to estimate gyroscope bias.

For the BROAD and Sassari datasets, we derive a ground truth for the gyroscope bias by averaging the gyroscope measurements during the rest phases at the beginning and at the end of each trial and linearly interpolating in between to account for slow bias instability.
For each \gls{IOE} algorithm, we calculate the root-mean-square over time of the residual bias norm, i.e., of the norm of the difference between the estimated bias and the true bias.
\autoref{fig:vqf_bias_estimation_plot} shows the achieved relative reduction of the gyroscope bias, the true bias norm, and the residual bias norm, averaged over all trials for each dataset.
Since the proposed VQF and the literature method VAC use rest detection for bias estimation, while the other algorithms do not, we also test the performance of all algorithms on a cut version of the \gls{BROAD} dataset, in which the initial and final rest phases were removed.

For the proposed algorithms VQF and OfflineVQF, we present the bias estimates obtained with the default parameters.
For the existing methods, we found that parameters that yield the best orientation estimation results often do not yield the best bias estimates.
We therefore optimized, separately for each dataset, all parameters of all literature methods (except MKF) across the search grid presented in \autoref{tab:vqf_algorithm_params}, such that the bias estimation error is minimized.
Despite this disparity, the proposed method clearly outperforms the bias estimation methods of all other \gls{IOE} algorithms.
Even though the results obtained with the cut \gls{BROAD} dataset are worse than for the datasets with long rest phases, the proposed method is still able to reduce the bias by \SI{39}{\percent}, while the best literature method only achieves a reduction of \SI{18}{\percent}.
As with orientation estimation, using the OfflineVQF variant further improves the accuracy in comparison to the real-time capable VQF algorithm.
MKF only achieves a slight reduction of gyroscope bias for the Sassari dataset, while for BROAD the bias norm increases compared to the original bias found in the measurement data.

\begin{figure*}[t]
\includegraphics{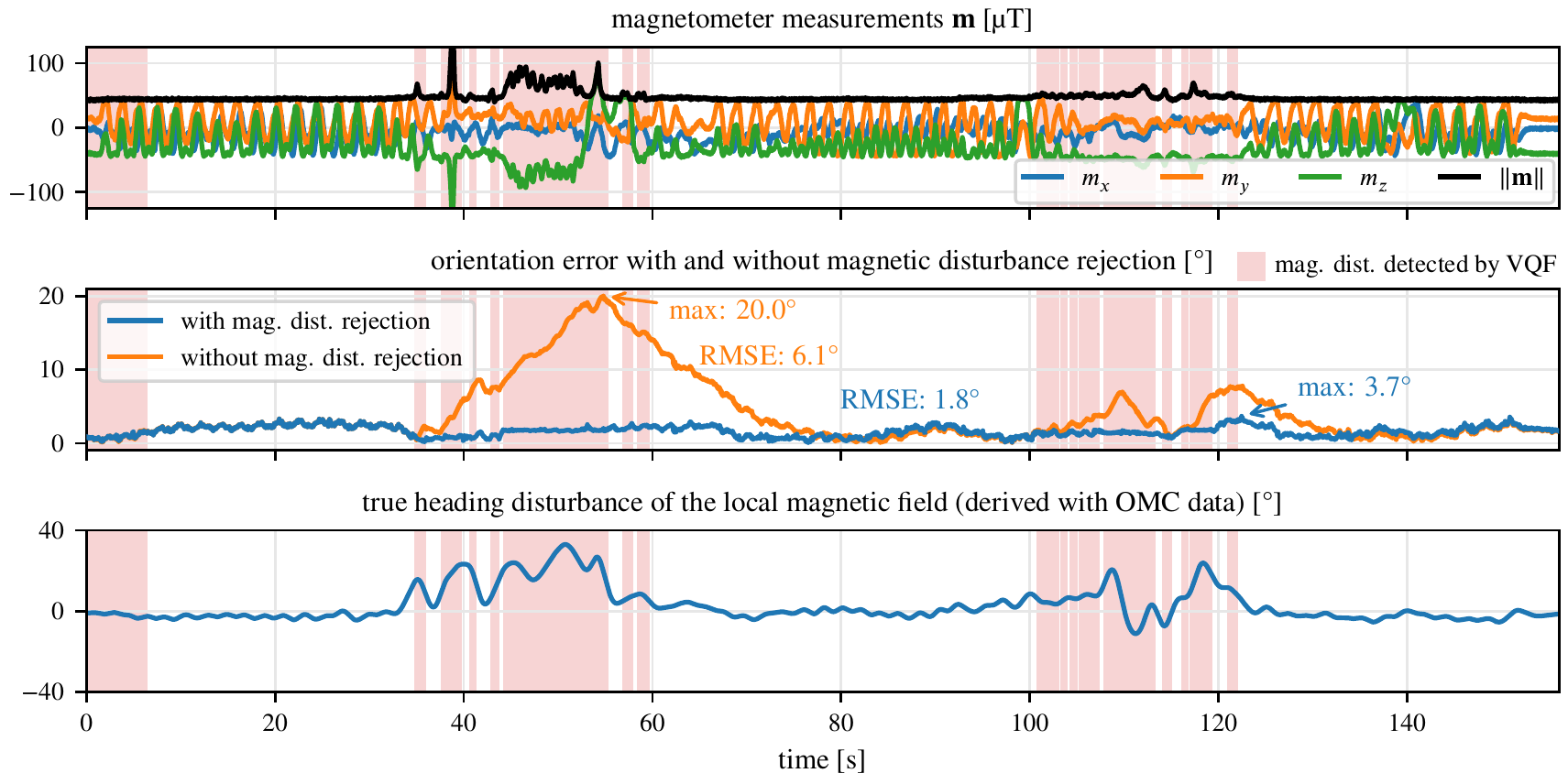}
\caption{Performance of the magnetic disturbance rejection method for trial 37 of the BROAD dataset (disturbed office environment). Whenever magnetic disturbances are detected by \autoref{alg:magdist} (light red background), the magnetometer-based correction is automatically disabled. In comparison to the same algorithm with disabled magnetic disturbance rejection, the RMSE is considerably reduced, and the large error peak of $\ang{20}$ is avoided.}
\label{fig:vqf_magdist_example_plot}
\end{figure*}

To illustrate how the bias estimation method works, \autoref{fig:vqf_bias_example_plot} shows the estimated bias, the estimation error, and the estimation uncertainty for an exemplary trial of the \gls{BROAD} dataset.
It can be seen that the bias estimation works as intended:
During the initial motion phase, the estimated bias slowly converges to the true bias.
Once rest is automatically detected, the alternative update is used, causing the estimate to rapidly converge.

In summary, both in a systematic comparison, as well as in an exemplary case study, the proposed gyroscope bias estimation method was found to work reliably and clearly outperformed existing bias estimation approaches.

\subsection{Magnetic Disturbance Rejection}

To further illustrate the difference in performance between the proposed BasicVQF and VQF algorithms, we now take a brief look at the performance of the magnetic disturbance detection and rejection method.
\autoref{fig:vqf_magdist_example_plot} shows the behavior of this extension on an example of the \gls{BROAD} dataset.
For the sake of evaluation, \gls{OMC} data was used to determine the true disturbance of the local magnetic field caused by ferromagnetic material and electric devices in the office environment.
This ground truth information shows that the detection is triggered whenever disturbances are present.
Without magnetic disturbance rejection, the orientation estimation error reaches a maximum of \ang{20.0}, and the \gls{RMSE} is \ang{6.1}.
In contrast, enabling magnetic disturbance rejection reduces the maximum error to just \ang{3.7} and the \gls{RMSE} to \ang{1.8}, which translates to an at least three times better accuracy.
This example demonstrates how the optional magnetic disturbance rejection can improve the reliability of 9D \gls{IOE} in real-world scenarios inside buildings, near ferromagnetic material and electric devices.

\subsection{Summary of the Results}\label{sec:result_summary}

The proposed VQF algorithm achieved an average \gls{RMSE} of \ang{2.9} for 9D \gls{IOE}, while the average errors obtained with state-of-the-art methods range from \ang{5.3} to \ang{16.7}.
For 6D \gls{IOE}, VQF attained an average \gls{RMSE} of \ang{1.1}, compared to \ang{2.4}--\ang{6.3} obtained with existing methods, and it achieved even \SI{17}{\percent} lower errors than a neural network that was trained on large portions of the benchmark data.
For the 13 characteristic trial groups of the \gls{BROAD} dataset, the proposed method outperforms even the best-performing literature method for all seven motion characteristics and for four out of six disturbance characteristics.
Furthermore, the gyroscope bias estimation of VQF clearly outperformed all existing state-of-the-art literature methods and compensated $\sim\SI{90}{\percent}$ of the bias.
Even for the challenging case without rest phases, the bias could still be reduced by $\sim$\SIrange[range-phrase=--,range-units=single]{40}{60}{\percent}, while the existing algorithms barely achieved any bias reduction.
For an exemplary case in a simulated office environment, the magnetic disturbance rejection algorithm was shown to achieve a five-fold reduction of the maximum orientation error.

Even the variant BasicVQF, without bias estimation and magnetic disturbance rejection, was shown to provide clearly more accurate 9D orientation estimates than all state-of-the-art methods.
For applications in which real-time capability is not required, the variant OfflineVQF can be used to further increase accuracy.

\section{Conclusions}

We proposed a novel \gls{IOE} algorithm that simultaneously performs 6D and 9D sensor fusion, estimates gyroscope bias, and performs magnetic disturbance detection and rejection.
An open-source implementation is provided in C++, Python, and Matlab, making it easy to use the algorithm.
We compared the proposed algorithm VQF with eight other \gls{IOE} algorithms and using a collection of six publicly available datasets that cover a wide range of motions, speeds, disturbances, and different sensor hardware.
As summarized in \autoref{sec:result_summary}, VQF consistently provided the best performance, both for 6D and 9D orientation estimation, as well as for gyroscope bias estimation, and it proved capable of magnetic disturbance rejection.

The proposed method provides a highly accurate out-of-the-box performance, which means that -- unlike existing literature methods -- VQF requires no parameter tuning for a vast range of motions and application scenarios.
For rare edge cases, the proposed method facilitates easy and intuitive tuning via the time constants \tauAcc{} and \tauMag{}.

The achieved improvements in ease of use and in orientation estimation accuracy are expected to change the way we use \gls{IOE} algorithms in practice and, thereby, to advance the broad field of inertial motion tracking since it enables more accurate IMU-based position and velocity estimation, joint angle estimation, and 3D visualization.
This, in turn, leads to improved performance in many existing application areas of miniature inertial sensor technology, and it likewise facilitates the applicability in novel application domains with increased accuracy demands.

Future work will focus on integrating continuous and automatic magnetometer calibration as well as on employing the VQF algorithm in various applications.

\printcredits

\bibliographystyle{cas-model2-names_NOSORT}

\bibliography{bibliography}

\appendix

\section{Details on the Basic Update Step}\label{app:vqf_basic_step}

The basic filter update given in \sref{Algorithm}{alg:basicvqf} consists of a prediction of the next orientation based on the gyroscope measurement, followed by a correction of the inclination based on the accelerometer measurement, and an (optional) correction of the heading based on the magnetometer measurement.

\subsection{Gyroscope Prediction}

Gyroscope prediction via strapdown integration is performed by multiplying the previous estimate with a quaternion based on the norm and direction of the measured angular rate:
\begin{align}
\quat{\imui(\tk)}{\inertiali(\tk)} &= \quat{\imui(\tk[-1])}{\inertiali(\tk[-1])} \quatmult \quataa{T_s\lVert\bm{\omega}\rVert}{\bm{\omega}}.\label{eq:gyrstrapdown}
\end{align}

The rotation due to the gyroscope measurement is composed of the true change of sensor orientation and an error, due to gyroscope bias, noise, and other measurement errors (e.g., scaling errors, nonlinearity, misalignment, and clipping).
This error can be regarded as a small drift in the $\inertiali$ frame, i.e., as $\quat{\inertiali(\tk[-1])}{\inertiali(\tk)}$, which can be shown via quaternion algebra (cf.\ \autoref{app:vqf_inertial_drift}).

\subsection{Accelerometer Correction}

The accelerometer measurements consist of the gravitational acceleration, change of velocity, as well as noise, bias, and other measurement errors.
Most existing methods \cite{madgwick2010efficient,mahony2008nonlinear,seel2017eliminating} interpret each single accelerometer sample as a 3D vector and use the angle between this vector and the expected vertical direction to derive the correction step.
In contrast, to better separate the gravitational acceleration from the other components of the measurement, we transform the measured accelerations into the almost-inertial frame $\inertiali$ and then apply a linear low-pass filter to each component.
The resulting signal provides a vertical reference in the  $\inertiali$ frame, which slowly drifts due to errors in gyroscope integration, as shown in \autoref{fig:vqf_accplot}.
As a low-pass filter, we use a second-order Butterworth filter.
The cutoff frequency $f_\mathrm{c,acc}$ of this filter defines the weight between gyroscope prediction and accelerometer correction.
To ensure a fast and robust convergence when the algorithm is initialized, we calculate the arithmetic mean for the first few samples instead of using the Butterworth filter.
Then, the filter state is initialized based on this mean value.

\begin{figure}[b]
\includegraphics{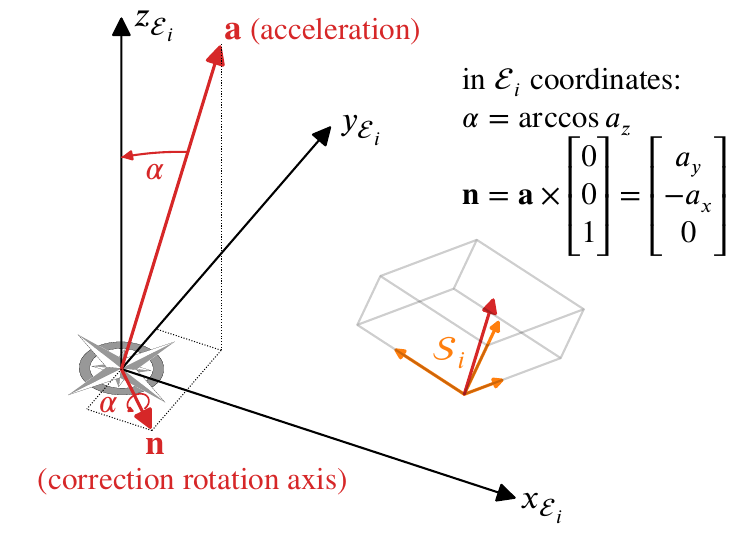}
\caption{Illustration of the inclination correction step based on the filtered accelerometer measurement. In global $\earthi$ coordinates, the filtered and normalized acceleration $\nvec{a}{}=\inlinevec{a_x,a_y,a_z}$ is expected to point in positive $z$-direction. This can be achieved by a correction rotation with angle $\arccos a_z$ and axis $\inlinevec{a_y,-a_x,0}$.}
\label{fig:vqf_acc_update}
\end{figure}

As illustrated in \autoref{fig:vqf_acc_update}, we can use this vertical reference to correct the inclination estimate.
If $\inlinevec{a_x,a_y,a_z}$ denotes the filtered and normalized acceleration measurement in the $\earthi$ frame, the shortest rotation for inclination correction has an angle of $\arccos a_z$ around the axis $\inlinevec{a_x,a_y,a_z} \times \inlinevec{0,0,1} = \inlinevec{a_y,-a_x,0}$.
The quaternion \nquat{corr} that corresponds to this rotation can be expressed without trigonometric functions as
\begin{align}
q_w &= \cos\left(\frac{\arccos a_z}{2}\right) = \sqrt{\frac{a_z + 1}{2}}\\
\nquat{corr} &= \begin{bmatrix}q_w & \frac{a_y}{2 q_w} & \frac{-a_x}{2 q_w} & 0\end{bmatrix}\tps.\label{eq:vqf_acc_corr_quat}
\end{align}
This quaternion is used to correct the estimate of the $\quat{\inertiali}{\earthi}$ quaternion, i.e.,
\begin{equation}
\quat{\inertiali(\tk)}{\earthi(\tk)} = \nquat{corr}(\tk) \quatmult \quat{\inertiali(\tk[-1])}{\earthi(\tk[-1])}.
\end{equation}
After the correction step, the low-pass filtered acceleration will perfectly point in upward direction, i.e., in $z$-direction of the \earthi{} and \earth{} frames.

\subsection{Magnetometer Correction}

\begin{figure}[b]
\includegraphics{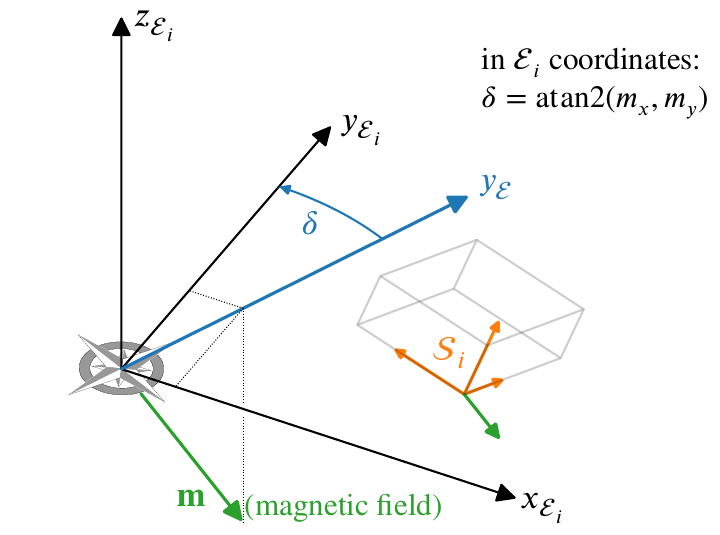}
\caption{Illustration of the heading correction step based on the magnetometer measurement. In many parts of the world, Earth's magnetic field is dominated by a vertical component, e.g., in Berlin, the magnetic field is pointing down with a dip angle of \ang{68}. An estimate of the magnetic north direction can be obtained by projecting the measured magnetic field into the horizontal plane.}
\label{fig:vqf_mag_update}
\end{figure}

If magnetometer measurements are given, we use them to correct the heading estimate.
As shown in \autoref{eq:vqf_state}, the heading is tracked via a scalar state $\delta_i(\tk)$ that represents the vertical rotation from the global $\earth$ frame (with the $y$-axis pointing north) to the $\earthi$ frame.
As illustrated in \autoref{fig:vqf_mag_update}, we can use the current magnetometer sample to derive a measurement $\delta_\mathrm{mag}(\tk)$ for this state by projecting the magnetic field vector into the horizontal plane.
The state is then corrected by a fixed fraction $k_\mathrm{mag}$ of the deviation between state and measurement, which corresponds to a first-order low-pass filter with exponential convergence.

The parameter $k_\mathrm{mag}$ defines the fusion weight between gyroscope prediction and magnetometer correction.
To ensure robust and fast convergence when the filter is initialized with the default value $\delta_i(t_0)=0$, we average the first measurements by choosing the filter weight $k_\mathrm{mag}$ as $1, \frac{1}{2}, \frac{1}{3}, \ldots$ during the first $\frac{1}{k_\mathrm{mag}}$ steps.

\subsection{Definition of Intuitive Fusion Weights}

The behavior of the algorithm can be influenced by two fusion weights $f_\mathrm{c,acc}$ and $k_\mathrm{mag}$, which we will now replace with a more intuitive parametrization by defining intuitive time constants $\tauAcc$ and $\tauMag$ that allow the user to influence the fusion weights between gyroscope prediction and accelerometer correction and between gyroscope prediction and magnetometer correction, respectively.
The parameters $f_\mathrm{c,acc}$ and $k_\mathrm{mag}$ that are internally used in the filter update step (\sref{Algorithm}{alg:basicvqf}) are automatically derived from those time constants.

In order to specify the gain $k_\mathrm{mag}$ of the magnetometer correction first-order exponential filter, we use the time constant $\tau=\frac{1}{2\pi f_\mathrm{c}}$ that is commonly used to characterize first-order systems and corresponds to the time needed for the step response to reach $1-e^{-1} \approx \SI{63.2}{\percent}$ of its final value.
The filter weight for the proportional update can be derived from this time constant as
\begin{equation}
k_\mathrm{mag} = 1 - \exp\left(-\frac{\Ts}{\tauMag}\right).
\end{equation}

\begin{figure}[b]
\includegraphics{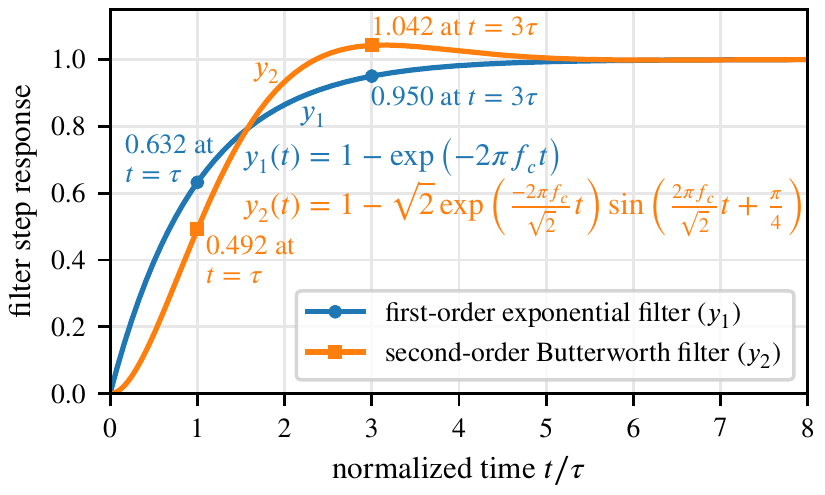}
\caption{Step response for first- and second-order low-pass filters. The time axis is normalized by dividing by the time constant $\tau$ of the proposed parametrization. Both filter outputs are roughly close to 0.5 at $t=\tau$ and converge up to a deviation of $\le\SI{5}{\percent}$ at $t=3\tau$.}
\label{fig:vqf_step_response_tau_plot}
\end{figure}

The second-order Butterworth filter used for accelerometer correction is characterized by the cutoff frequency $f_\mathrm{c,acc}$.
In order to obtain a parametrization that is similar to the parametrization of the magnetometer correction, we use a time constant that corresponds to the undampened part of the step response, i.e., $\tauAcc=\frac{\sqrt{2}}{2\pi f_\mathrm{c,acc}}$.
The cutoff frequency used to determine the Butterworth filter coefficients is then given as
\begin{equation}
f_\mathrm{c,acc}=\frac{\sqrt{2}}{2\pi\tauAcc}.
\end{equation}
See \autoref{fig:vqf_step_response_tau_plot} for a comparison of the step responses of both filter types in relation to the time constant $\tau$.

This mapping is used to derive the internal parameters $f_\mathrm{c,acc}$ and $k_\mathrm{mag}$ from the user-specified time constants $\tauAcc$ and $\tauMag$.

\section{Gyroscope Integration Errors Cause Drift of Almost-Inertial Frame}\label{app:vqf_inertial_drift}

Every gyroscope prediction step causes a small error, due to gyroscope bias, noise, and other measurement errors.
Without correction by accelerometers and magnetometers, those errors will add up and lead to drift in the orientation estimates.
We can show that this error can be regarded as a small drift in the $\inertiali$ frame, i.e., all rotation that happens in the gyroscope prediction and that is not the true change of sensor orientation can mathematically be expressed as a rotation $\quat{\inertiali(\tk[-1])}{\inertiali(\tk)}$.

The gyroscope prediction step consists of multiplication of the previous estimate with an update quaternion based on the measured angular rate:
\begin{equation}
\quat{\imui(\tk)}{\inertiali(\tk)} = \quat{\imui(\tk[-1])}{\inertiali(\tk[-1])} \quatmult \quataa{T_s\lVert\bm{\omega}\rVert}{\bm{\omega}}.
\end{equation}
This update quaternion can be expressed as the true change in sensor orientation multiplied with a small error quaternion \nquat{e}, i.e.,
\begin{equation}
\quataa{T_s\lVert\bm{\omega}\rVert}{\bm{\omega}} = \quatimuimu{_i(\tk)}{_i(\tk[-1])} \quatmult \nquat{e}.
\end{equation}
We can transform the error rotation \nquat{e} to any frame, here $\inertiali(\tk[-1])$:
\begin{equation}
\nquat{e} = \quat{\inertiali(\tk[-1])}{\imui(\tk)} \quatmult \nquat[{\inertiali(\tk[-1])}]{e} \quatmult \quat{\inertiali(\tk[-1])}{\imui(\tk)}\quatinv.
\end{equation}
When putting this into the prediction step, we obtain
{\setlength{\mathindent}{0cm}
\begin{align}
\quat{\imui(\tk)}{\inertiali(\tk)} &= \quat{\imui(\tk[-1])}{\inertiali(\tk[-1])} \quatmult \quataa{T_s\lVert\bm{\omega}\rVert}{\bm{\omega}}\\
&= \quat{\imui(\tk[-1])}{\inertiali(\tk[-1])} \quatmult \quatimuimu{_i(\tk)}{_i(\tk[-1])} \quatmult \nquat{e}\\
&= \quat{\imui(\tk)}{\inertiali(\tk[-1])} \quatmult \quat{\inertiali(\tk[-1])}{\imui(\tk)} \quatmult \nquat[{\inertiali(\tk[-1])}]{e} \quatmult \quat{\inertiali(\tk[-1])}{\imui(\tk)}\quatinv\\
&= \underbrace{\nquat[{\inertiali(\tk[-1])}]{e}}_{\quat{\inertiali(\tk[-1])}{\inertiali(\tk)}} \quatmult \quat{\imui(\tk)}{\inertiali(\tk[-1])}
\end{align}}

Therefore, expressed in the almost-inertial frame $\inertiali(\tk[-1])$, the gyroscope prediction error quaternion \nquat{e} corresponds to the drift rotation $\quat{\inertiali(\tk[-1])}{\inertiali(\tk)}$ of the almost-inertial frame $\inertiali$.

\section{Details on Gyroscope Bias Estimation}\label{app:vqf_bias_est}

The full algorithm for gyroscope bias estimation is given in \autoref{alg:biasest}.

\begin{algorithm}[htb]
\caption{Gyroscope Bias Estimation}
\label{alg:biasest}
\begin{algorithmic}[1]
    \Procedure{RestDetection}{$\gyr{},\nvec{a}{}$}
        \State $T_\mathrm{rest} \gets T_\mathrm{rest} + \Ts$
        \For{each measurement $\gyr{},\nvec{a}{}$}
            \State low-pass filter the measurement with $\tau=\SI{0.5}{\second}$
            \If{absolute difference between current and filtered measurement value is above threshold for any component}
                \State $T_\mathrm{rest} \gets 0$
            \EndIf
        \EndFor
        \If{$T_\mathrm{rest} \ge \SI{1.5}{\second}$}
            \State rest detected
        \Else
            \State movement detected
        \EndIf
    \EndProcedure
    \Procedure{InitializeKalmanFilter}{}
        \State $\nvec{\hat{b}}{} \gets \inlinevec{0,0,0}$\Comment{Gyroscope bias estimate}
        \State $\nvec{P}{} \gets (\SI{0.5}{\dps})^2 \mathbf{I}_{3\times3}$\Comment{Covariance matrix}
        \State $v \gets (\SI{0.1}{\dps})^2 \Ts (\SI{100}{\second})^{-1}$\Comment{System noise}
        \State $w_\mathrm{motion} \gets (\SI{0.1}{\dps})^4 v^{-1} + (\SI{0.1}{\dps})^2$\Statex\Comment{Motion update variance}
        \State $w_\mathrm{rest} \gets (\SI{0.03}{\dps})^4 v^{-1} + (\SI{0.03}{\dps})^2$\Statex\Comment{Rest update variance}
    \EndProcedure
    \Procedure{BiasEstimationStep}{$\quat{\inertiali}{\earthi},a_x,a_y,a_z$}
        \State $\nvec{R}{} \gets$ rotation matrix corresponding to $\quat{\inertiali}{\earthi}$
        \State $\nvec{R}{LP} \gets$ low-pass filter $\nvec{R}{}$ with $\tau = \tau_\mathrm{acc}$
        \State $\nvec{\hat{b}}{\earthi,LP} \gets$ low-pass filter $\nvec{R}{}\nvec{\hat{b}}{}$ with $\tau = \tau_\mathrm{acc}$
        \If{rest detected}
            \State $\nvec{y}{} \gets \nvec{\hat{b}}{}$
            \State $\nvec{C}{} \gets \mathbf{I}_{3\times3}$
            \State $\nvec{W}{} \gets w_\mathrm{rest} \inlinevec{1,1,1}$
        \Else%
            \State $\nvec{y}{} \gets \Ts^{-1} \inlinevec{a_y,-a_x,0} + \operatorname{diag}(1,1,0) \nvec{\hat{b}}{\earthi,LP}$
            \State $\nvec{C}{} \gets \nvec{R}{LP}$
            \State $\nvec{W}{} \gets w_\mathrm{motion}\inlinevec{1,1,\frac{1}{0.0001}}$
        \EndIf
        \State $\nvec{P}{} \gets \nvec{P}{} + v \inlinevec{1,1,1}$ \Comment{Kalman filter update}
        \State $\nvec{K}{} \gets \nvec{P}{} \nvec{C}{}\tps (\nvec{W}{} + \nvec{C}{} \nvec{P}{} \nvec{C}{}\tps)^{-1}$
        \State $\nvec{\hat{b}}{} \gets \nvec{\hat{b}}{} + \nvec{K}{} \operatorname{clip}(\nvec{y}{} - \nvec{C}{} \nvec{\hat{b}}{}, \SI{-2}{\dps}, \SI{2}{\dps})$ \Statex\Comment{Limit disagreement to \SI{2}{\dps}}
        \State $\nvec{P}{} \gets \nvec{P}{} - \nvec{K}{} \nvec{C}{} \nvec{P}{}$
        \State $\nvec{\hat{b}}{} \gets \operatorname{clip}(\nvec{\hat{b}}{}, \SI{-2}{\dps}, \SI{2}{\dps})$\Statex\Comment{Limit bias estimate to \SI{2}{\dps}}
    \EndProcedure
\end{algorithmic}
\end{algorithm}

To detect whether the \gls{IMU} is at rest (procedure \textproc{RestDetection} in \autoref{alg:biasest}), we first filter each component of the gyroscope and accelerometer measurements with a second-order Butterworth filter and a time constant of $\tau=\SI{0.5}{\second}$.
Note that, unlike the low-pass filter for the acceleration used for inclination correction, we apply the filter directly in the sensor frame.
We then calculate the Euclidean norm of the deviation between the current measurement and the filtered measurement.
Rest is detected if, in the last $\SI{1.5}{\second}$, the gyroscope and accelerometer deviations are always less than $\SI{2}{\dps}$ and $\SI{0.5}{\mpss}$, respectively.
Note that we deliberately do not use magnetometer measurements for rest detection since we found the rest detection to be very reliable when only using gyroscope and accelerometer measurements, whereas using magnetometers did not add additional value.

To estimate the gyroscope bias during rest and motion, we employ a Kalman filter \cite{kalman1960new,welch2006introduction}, with the gyroscope bias as state and a time-dependent output matrix:
{\setlength{\mathindent}{0cm}
\begin{align}
\mathbf{b}(\tk) &= \mathbf{b}(\tk[-1]) + \mathbf{v}(\tk),& \mathbf{v}(\tk) &\sim \mathcal{N}(0, \mathbf{V})&&\\
\mathbf{y}(\tk) &= \mathbf{C}(\tk) \mathbf{b}(\tk) + \mathbf{w}(\tk),& \mathbf{w}(\tk) &\sim \mathcal{N}(0, \mathbf{W}(\tk)).
\end{align}}

During rest, we use the low-pass filtered gyroscope readings $\gyr{LP}$ (which we calculated in the rest detection procedure) as a direct measurement of the bias, i.e., $\mathbf{C}(\tk)=\mathbf{I}_{3\times3}$ and $\mathbf{y}(\tk) =\gyr{LP}(\tk)$.
Because the \gls{IMU} is at rest and gyroscope readings are already filtered, we can assign a comparatively large weight (i.e., a small covariance) to this measurement update and achieve fast convergence for the bias estimate $\mathbf{\hat{b}}(\tk)$.

\begin{figure}[b]
  \includegraphics{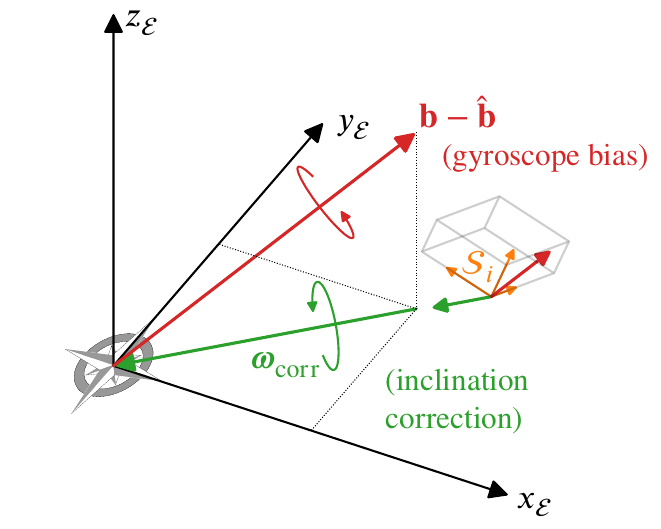}
  \caption{Illustration of the principle behind gyroscope bias estimation from the inclination correction step. In the steady state, the correction angular rate $\gyr{corr}$ that corresponds to the accelerometer-based correction is the (negative) horizontal projection of the remaining gyroscope bias $\mathbf{b}-\hat{\mathbf{b}}$.}
  \label{fig:vqf_bias_projection}
\end{figure}

During motion, we estimate the gyroscope bias from the inclination correction steps.
At every time step, the new error due to gyroscope bias is a local rotation (i.e., in $\imui$) with the rotation vector $T_s (\mathbf{b}-\mathbf{\hat{b}})$, and the inclination correction is a global rotation (i.e., in $\earthi$) with a horizontal rotation vector $\mathbf{c} = \inlinevec{c_x,c_y,0}$.
In ideal conditions (i.e., in a steady state and without noise or other errors), the correction rotation will exactly compensate the inclination portion of the bias rotation.
As illustrated in \autoref{fig:vqf_bias_projection}, in this case, the correction is the inverse of the horizontal projection of the bias rotation.

With $\mathbf{R}$ being the rotation matrix corresponding to \quat{\imui}{\earthi}, we can express this as
\begin{equation}
\mathbf{R} (\mathbf{b}-\mathbf{\hat{b}}) = -\frac{1}{\Ts}\begin{bmatrix}c_x\\c_y\\{*}\end{bmatrix},
\end{equation}
where the star indicates that the gyroscope bias in the current vertical direction is not observable by accelerometer measurements.

To achieve slow forgetting of the bias for the special case in which the same axis is vertical for a long time, we set the corresponding measurement to zero, with a significantly larger variance to slow down convergence.

With an output matrix of $\mathbf{C}_k = \mathbf{R} =: (r_{ij})$ and the correction vector $\inlinevec{a_y,-a_x,0}$, we obtain the following measurement equation:
\begin{equation}
\mathbf{y}_k = \begin{bmatrix}
-\frac{1}{\Ts} a_y + r_{11} \hat{b}_x + r_{12} \hat{b}_y + r_{13} \hat{b}_z\\
\phantom{-}\frac{1}{\Ts} a_x + r_{21} \hat{b}_x + r_{22} \hat{b}_y + r_{23} \hat{b}_z\\
0
\end{bmatrix}.\label{eq:vqf_motionbias_measurement}
\end{equation}
This measurement equation does not take into account that the measured accelerations are low-pass filtered in the $\earthi$ frame.
The accuracy and robustness of the bias estimation can be significantly improved by low-pass filtering the components of $\mathbf{R}$ and $\mathbf{R}\hat{\mathbf{b}}$ with the same low-pass filter used for the accelerometer measurements in the $\inertiali$ frame.
See \autoref{app:vqf_filter_rotmat} for a detailed derivation of this full measurement equation.

The tuning parameters of a Kalman filter are the initial covariance, the covariance of the system noise, and the covariance of the measurement noise.
Since the real values of those covariances are hard to obtain and would furthermore depend on the sampling rate, we employ a parametrization that ensures the following properties:
\begin{enumerate}
\item The initial estimation uncertainty is $\sigma_\mathrm{init} = \SI{0.5}{\dps}$.
\item During motion, the uncertainty converges to $\sigma_\mathrm{motion}=\SI{0.1}{\dps}$ (for the non-vertical axes).
\item During rest, the uncertainty converges to $\sigma_\mathrm{rest}=\SI{0.03}{\dps}$.
\item Without updates, the estimation uncertainty increases from 0 to $\SI{0.1}{\dps}$ in the forgetting time $t_\mathrm{forget} = \SI{100}{\second}$.
\end{enumerate}
How those parameters translate to the covariances internally used by the Kalman filter is specified in \autoref{alg:biasest} and further explained in \autoref{app:vqf_bias_params}.
Note that the absolute values of the provided parameters are arbitrary and chosen to facilitate an intuitive understanding of the estimation uncertainty.
For the behavior of the Kalman filter, only the relation between the parameters is relevant.

\section{Measurement for Motion Bias Estimation}\label{app:vqf_filter_rotmat}

The approach for gyroscope bias estimation during motion, as shown in \autoref{fig:vqf_bias_projection}, is based on the assumption that the current inclination correction step corresponds to the gyroscope bias that is transformed to the global frame using the current sensor orientation.
This works well if the sensor orientation does not change much.
In reality, due to the low-pass filter used for the acceleration, the inclination correction corresponds to the bias-induced rotations from the last few seconds, taking the respective sensor orientations into account.
In the following, we show that this effect is well-described by applying the same low-pass filter to the elements of the rotation matrices $\mathbf{R}$ corresponding to $\quat{\imui}{\inertiali}$, and also to the rotated bias estimates $\mathbf{R}\hat{\mathbf{b}}$.

To simplify the notation, we introduce the rotation operator:
\begin{equation}
\rot{\nquat{}}{\nvec{v}{}} := \nquat{} \quatmult \nvec{v}{} \quatmult \nquat{} \quatinv.
\end{equation}

Assume that the accelerometer measurements are always a perfect vertical vector ($\nvec[\earthtrue]{a\mathit{(\tk)}}{} = \nvec[\earthtrue]{v}{} = \inlinevec{0,0,1}$), i.e. there are no disturbances or measurement errors (unit length is only used to simplify the notation).
In the orientation estimation update step, those accelerometer measurements are transformed into the $\inertiali(\tk)$ frame and then low-pass filtered with an \gls{IIR} filter with impulse response $b_n$, i.e.,
\begin{align}
\coordtrafobrackets{\inertiali(\tk)}{\nvec{a}{LP}(\tk)} &= \sum\limits_{n=0}^\infty b_n \left[
\rot{\quat{\earthtrue}{\inertiali(\tk[-n])}}{
\nvec[\earthtrue]{v}{}}
\right]\\
&= \sum\limits_{n=0}^\infty b_n
\nvec[{\inertiali(\tk[-n])}]{v}{}.
\end{align}
One time step earlier, the filter output is
\begin{align}
\coordtrafobrackets{\inertiali(\tk[-1])}{\nvec{a}{LP}(\tk[-1])}
&= \sum\limits_{n=0}^\infty b_n \nvec[{\inertiali(\tk[-1-n])}]{v}{}\\
&= \sum\limits_{n=0}^\infty b_n
\rot{\quat{\inertiali(\tk[-n])}{\inertiali(\tk[-1-n])}}{\nvec[{\inertiali(\tk[-n])}]{v}{}}.
\end{align}
From Rodrigues' rotation formula follows that for a small rotation $\theta$ around axis $\nvec{k}{}$,
\begin{equation}
\nvec{p}{rot} \approx \nvec{p}{} + \theta \nvec{k}{} \times \nvec{p}{}.
\end{equation}
We use this approximation for the rotation of the $\inertiali$ frame due to gyroscope bias in one sample step $\quat{\inertiali(\tk[-1-n])}{\inertiali(\tk[-n])}$ (cf.\ \autoref{app:vqf_inertial_drift}).
Note that this rotation is only caused by the uncorrected part of the gyroscope bias, i.e., by $\mathbf{b}'(\tk) := \nvec{b}{} - \mathbf{\hat{b}}(\tk)$.
Therefore, we obtain
{\setlength{\mathindent}{0cm}
\begin{align}
&\coordtrafobrackets{\inertiali(\tk[-1])}{\nvec{a}{LP}(\tk[-1])}\\&\approx \sum\limits_{n=0}^\infty b_n \left(
\nvec[{\inertiali(\tk[-n])}]{v}{}
- \Ts \coordtrafobrackets{{\inertiali(\tk[-n])}}{\mathbf{b}'(\tk[-n])}
\times
\nvec[{\inertiali(\tk[-n])}]{v}{}
\right).
\end{align}}

The difference between two consecutive filtered accelerations is then
{\setlength{\mathindent}{0cm}
\begin{align}
&\coordtrafobrackets{\inertiali(\tk)}{\nvec{a}{LP}(\tk)} - \coordtrafobrackets{\inertiali(\tk[-1])}{\nvec{a}{LP}(\tk[-1])} \\&\approx  \Ts \sum\limits_{n=0}^\infty b_n \left(
\coordtrafobrackets{{\inertiali(\tk[-n])}}{\mathbf{b}'(\tk[-n])}
\times
\nvec[{\inertiali(\tk[-n])}]{v}{}
\right)\\
&= \Ts \sum\limits_{n=0}^\infty b_n \left(
\rot{\quat{\imui(\tk[-n])}{\inertiali(\tk[-n])}}{\nvec{b}{}'(\mathit{\tk[-n]})}
\times
\nvec[{\inertiali(\tk[-n])}]{v}{}
\right).
\end{align}}
Now, we express this difference in the frame $\earthi(\tk[-1])$ that is used to perform the inclination correction step, assuming that $\earthi$ and $\inertiali$ do not change much over the duration that is relevant for the filter (since the influence of bias and the inclination correction is limited during short time spans), and assuming that the true vertical axis is approximately $\inlinevec{0,0,1}$ in the $\earthi(\tk[-1])$ frame.
{\setlength{\mathindent}{0cm}
\begin{align}
&\coordtrafobrackets{\earthi(\tk[-1])}{\nvec{a}{LP}(\tk)} - \coordtrafobrackets{\earthi(\tk[-1])}{\nvec{a}{LP}(\tk[-1])} \\&\approx \Ts \sum\limits_{n=0}^\infty b_n \left[
\rot{\quat{\imui(\tk[-n])}{\earthi(\tk[-1-n])}}{\nvec{b}{}'(\tk[-n])}
\times
\begin{bmatrix}0\\0\\1 \end{bmatrix}
\right]\\
&= \Ts \left[\sum\limits_{n=0}^\infty b_n
\rot{\quat{\imui(\tk[-n])}{\earthi(\tk[-1-n])}}{\nvec{b}{}'(\tk[-n])}
\right]
\times
\begin{bmatrix}0\\0\\1 \end{bmatrix}
\end{align}}

Expressing the rotation by $\quat{\imui(\tk[-n])}{\earthi(\tk[-n-1])}$ with a rotation matrix $\mathbf{R}(\tk[-n])$ and introducing an $\operatorname{LPF}$ operator to simplify the notation of the low-pass filter yields
{\setlength{\mathindent}{0cm}
\begin{align}
&\coordtrafobrackets{\earthi(\tk[-1])}{\nvec{a}{LP}(\tk)} - \coordtrafobrackets{\earthi(\tk[-1])}{\nvec{a}{LP}(\tk[-1])} \\&= \Ts \left[\sum\limits_{n=0}^\infty b_n
\left(
\mathbf{R}(\tk[-n])
\nvec{b}{}'(\tk[-n])
\right)
\right]
\times
\begin{bmatrix}0\\0\\1 \end{bmatrix}\\
&= \Ts \operatorname{LPF}\left(\mathbf{R}(\tk) \nvec{b}{}'(\tk)\right)
\times
\begin{bmatrix}0\\0\\1 \end{bmatrix}.
\end{align}}
Since $\coordtrafobrackets{\earthi(\tk[-1])}{\nvec{a}{LP}(\tk[-1])}$ is always $\inlinevec{0,0,1}$ as the result of the previous inclination correction step, we get
{\setlength{\mathindent}{0cm}
\begin{align}
\coordtrafobrackets{\earthi(\tk[-1])}{\nvec{a}{LP}(\tk)} &= \begin{bmatrix}0\\0\\1 \end{bmatrix}
+ \Ts \operatorname{LPF}\left(\mathbf{R}(\tk) \nvec{b}{}'(\tk)\right)
\times
\begin{bmatrix}0\\0\\1 \end{bmatrix}.
\end{align}}
The inclination correction rotation vector $\nvec{c}{}(\tk)$, i.e., expressing the correction quaternion $\nquat{corr}$ from \autoref{eq:vqf_acc_corr_quat} as a rotation vector, is (neglecting the small change of the norm of $\coordtrafobrackets{\earthi(\tk[-1])}{\nvec{a}{LP}(\tk)}$)
{\setlength{\mathindent}{0cm}
\begin{align}
\nvec{c}{}(\tk) &= \coordtrafobrackets{\earthi(\tk)}{\nvec{a}{LP}(\tk[-1])} \times \begin{bmatrix}0\\0\\1 \end{bmatrix}\\
&= \left(
\begin{bmatrix}0\\0\\1 \end{bmatrix} + \Ts \operatorname{LPF}\left(\mathbf{R}(\tk) \nvec{b}{}'(\tk)\right)
\times
\begin{bmatrix}0\\0\\1 \end{bmatrix}
\right)
\times \begin{bmatrix}0\\0\\1 \end{bmatrix}\\
&= -\operatorname{diag}(1,1,0) \Ts \operatorname{LPF}\left(\mathbf{R}(\tk) \nvec{b}{}'(\tk)\right)\\
&= -\operatorname{diag}(1,1,0) \Ts \left(\operatorname{LPF}(\mathbf{R}(\tk))\nvec{b}{} - \operatorname{LPF}(\mathbf{R}(\tk)\mathbf{\hat{b}}(\tk))\right).
\end{align}}

Therefore, when choosing the system output matrix as $\mathbf{C}(\tk) = \operatorname{LPF}(\mathbf{R}(\tk))$, we obtain the following measurement in the horizontal plane:
{\setlength{\mathindent}{0cm}
\begin{align}
\operatorname{diag}(1,1,0) \mathbf{y}(\tk) &= - \frac{1}{\Ts} \nvec{c}{}(\tk) + \operatorname{diag}(1,1,0) \operatorname{LPF}(\mathbf{R}(\tk)\mathbf{\hat{b}}(\tk)).
\end{align}}

\section{Parametrization of the Bias Estimation Method}\label{app:vqf_bias_params}

Gyroscope bias is estimated using the system model
\begin{align}
\mathbf{b}_k &= \mathbf{b}_{k-1} + \mathbf{v}_k,& \mathbf{v}_k &\sim \mathcal{N}(0, \mathbf{V}),&&\\
\mathbf{y}_k &= \mathbf{C}_k \mathbf{b}_k + \mathbf{w}_k,& \mathbf{w}_k &\sim \mathcal{N}(0, \mathbf{W}_k),&&
\end{align}
where the index $k$ denotes sampling at $\tk$, and the standard Kalman filter update equations for the estimated state $\hat{\mathbf{b}}_k$
\begin{align}
\mathbf{P}_k^- &= \mathbf{P}^+_{k-1} + \mathbf{V}\\
\mathbf{K}_k &= \mathbf{P}_k^- \mathbf{C}_k\tps (\mathbf{W}_k + \mathbf{C}_k \mathbf{P}_k^- \mathbf{C}_k\tps)^{-1}\\
\hat{\mathbf{b}}_k &= \hat{\mathbf{b}}_{k-1} + \mathbf{K}_k (\mathbf{y}_k - \mathbf{C}_k \hat{\mathbf{b}}_{k-1})\\
\mathbf{P}_k^+ &= \mathbf{P}_k^- - \mathbf{K}_k \mathbf{C}_k \mathbf{P}_k^-.
\end{align}

The tuning parameters are the initial covariance $\mathbf{P}^+_{0}$, the variance of the system noise $\mathbf{V}$, and the variance of measurement noise $\mathbf{W}(\tk)$.
In the following, we derive an intuitive parametrization for those values that is independent of the sampling frequency.
Note that scaling all parameters of the Kalman filter with the same value will not change the system behavior.
For heuristically determined parameters, the actual quantities are therefore arbitrary.
However, a good parametrization still helps to make the behavior of the algorithm understandable and facilitates tuning.

A fixed amount of variance, the variance of the system noise $\mathbf{V}$, is added to the covariance matrix in every update step.
To be independent of the sampling frequency, scaling with the sampling time $\Ts$ is necessary.
To facilitate interpretation of the value as a \emph{forgetting time}, we parametrize the system noise by the time needed for the standard deviation of the estimation uncertainty to increase from 0 to $\SI{0.1}{\dps}$ in the absence of measurements, i.e.,
\begin{equation}
\mathbf{V} =  (\SI{0.1}{\dps})^2 \frac{\Ts}{t_\mathrm{forget}} \mathbf{I}_{3\times3}.
\end{equation}

We use the initial estimation uncertainty, i.e., the standard deviation  $\sigma_\mathrm{init}$, to initialize the covariance matrix:
\begin{equation}
\mathbf{P}^+_{0} = \sigma_\mathrm{init}^2 \mathbf{I}_{3\times3}.
\end{equation}

For the rest and motion updates, we provide the uncertainties $\sigma_\mathrm{rest}$ and $\sigma_\mathrm{motion}$ to which the estimate will eventually converge when the respective filter update is active instead of specifying the variance of the motion and rest update measurements directly.
This ensures independence of the sampling rates and makes the parameters easy to compare to the initial standard deviation.
The relation to the measurement variance $w_\mathrm{rest/motion}$ is given by
\begin{equation}
w_\mathrm{rest/motion} = \frac{\sigma_\mathrm{rest/motion}^4}{v} + \sigma_\mathrm{rest/motion}^2.\label{eq:vqf_measurement_noise_from_sigma}
\end{equation}

\begin{algorithm}[H]
\caption{Magnetic Disturbance Rejection}
\label{alg:magdist}
\begin{algorithmic}[1]
    \Procedure{MagDistDetection}{$\nvec{m}{}, \quat{\imui}{\earthi}$}
        \State $n \gets \lVert\nvec{m}{}\rVert$ \Comment{Norm of magnetic field}
        \State $\theta \gets -\arcsin\left(\inlinerowvec{0,0,1}\left(\quat{\imui}{\earthi}\quatmult\nvec{m}{}\quatmult\quat{\imui}{\earthi}\quatinv\right) n^{-1}\right)$\Statex\Comment{Dip angle}
        \State low-pass filter $n$ and $\theta$ with $\tau=\SI{0.05}{\second}$
        \If{$|n-n_\mathrm{ref}|<0.1n_\mathrm{ref}$ and $|\theta-\theta_\mathrm{ref}| < \ang{10}$}
            \State $T_\mathrm{undist} \gets T_\mathrm{undist}+\Ts$
            \If{$T_\mathrm{undist} \ge \SI{0.5}{\second}$}
                \State disturbed $\gets$ false
                \State $n_\mathrm{ref}\gets k_\mathrm{ref} \left(n - n_\mathrm{ref}\right)$
                \State $\theta_\mathrm{ref}\gets k_\mathrm{ref} \left(\theta - \theta_\mathrm{ref}\right)$\Statex\Comment{Track slow changes of norm and dip}
            \Else
                \State disturbed $\gets$ true
                \State $T_\mathrm{undist} \gets 0$
            \EndIf
        \EndIf
    \EndProcedure
    \Procedure{NewMagFieldAcceptance}{$n, \theta, |\gyr{}|$}
        \If{$|n-n_\mathrm{cand}|<0.1n_\mathrm{cand}$ and $|\theta-\theta_\mathrm{cand}| < \ang{10}$}
            \If{$\lVert\gyr{}\rVert \ge \SI{20}{\dps}$}\Statex\Comment{Only count the time if there is movement}
                \State $T_\mathrm{cand} \gets T_\mathrm{cand} + \Ts$
            \EndIf
            \State $n_\mathrm{cand}\gets k_\mathrm{ref} \left(n - n_\mathrm{cand}\right)$
            \State $\theta_\mathrm{cand}\gets k_\mathrm{ref} \left(\theta - \theta_\mathrm{cand}\right)$
            \If{disturbed and $T_\mathrm{cand} \ge \SI{20}{\second}$}\Statex\Comment{Accept candidate as new reference}
                \State disturbed $\gets$ false
                \State $n_\mathrm{ref} \gets n_\mathrm{cand}$
                \State $\theta_\mathrm{ref} \gets \theta_\mathrm{cand}$
            \EndIf
        \Else\Comment{Reset candidate to current value}
            \State $T_\mathrm{cand} \gets 0$
            \State $n_\mathrm{cand} \gets n$
            \State $\theta_\mathrm{cand} \gets \theta$
        \EndIf
    \EndProcedure
    \Procedure{MagDistRejection}{}
        \If{disturbed}
            \If{$T_\mathrm{reject} < \SI{60}{\second}$}
                \State $T_\mathrm{reject} \gets T_\mathrm{reject}+\Ts$
                \State do not perform heading correction
            \Else
                \State perform heading correction with $\frac{1}{2}k_\mathrm{mag}$
            \EndIf
        \Else
            \State $T_\mathrm{reject} \gets \operatorname{max}(T_\mathrm{reject}-2\Ts, 0)$
            \State perform heading correction with $k_\mathrm{mag}$
        \EndIf
    \EndProcedure
\end{algorithmic}
\end{algorithm}

\begin{algorithm*}[b]
\caption{Offline Orientation Estimation Algorithm}
\label{alg:vqf_offline}
\begin{algorithmic}[1]
    \Procedure{OfflineVQF}{$\gyr{}[t_{1:N}],\nvec{a}{}[t_{1:N}],\nvec{m}{}[t_{1:N}]$}
        \State $\ldots,\mathrm{dist}_1,\hat{\mathbf{b}}_1,\mathbf{P}_1 \gets \operatorname{VQF}(\gyr{}[t_{1:N}],\nvec{a}{}[t_{1:N}],\nvec{m}{}[t_{1:N}])$ \Comment{Run real-time filter in forward direction}
        \State $\ldots,\mathrm{dist}_2,\hat{\mathbf{b}}_2,\mathbf{P}_2 \gets \operatorname{VQF}(-\gyr{}[t_{N:1}],\nvec{a}{}[t_{N:1}],\nvec{m}{}[t_{N:1}])$ \Comment{Run real-time filter in backward direction}
        \State $\mathrm{dist}[t_{1:N}] \gets \mathrm{dist}_1[t_{1:N}] \land \mathrm{dist}_2[t_{1:N}] $ \Comment{Regard magnetic field as disturbed if both runs detected disturbances}
        \State $\hat{\mathbf{b}}[t_{1:N}] \gets (\mathbf{P}_1^{-1}+\mathbf{P}_2^{-1})^{-1}(\mathbf{P}_1^{-1} \hat{\mathbf{b}}_1 - \mathbf{P}_2^{-1} \hat{\mathbf{b}}_2)$\Comment{Average bias estimates of both filter runs via covariance}
        \State $\quat{\imui}{\inertiali}[t_{1:N}] \gets \operatorname{integrateGyr}(\gyr{}[t_{1:N}] - \hat{\mathbf{b}}[t_{1:N})$ \Comment{Perform gyroscope strapdown integration}
        \State $\nvec[\inertiali]{a}{}[t_{1:N}] \gets \quat{\imui}{\inertiali}[t_{1:N}]\quatmult\nvec{a}{}[t_{1:N}]\quatmult\quat{\imui}{\inertiali}[t_{1:N}]\quatinv$\Comment{Transform acceleration into $\inertiali$ frame}
        \State $\nvec[\inertiali]{a}{LP}[t_{1:N}] \gets \operatorname{filtfiltLPF}(\nvec[\inertiali]{a}{}[t_{1:N}], \tau=\tauAcc)$ \Comment{Forward-backward low-pass filtering}
        \State $\quat{\inertiali}{\earthi}[t_{1:N}] \gets$ perform inclination correction based on $\nvec[\inertiali]{a}{LP}[t_{1:N}]$
        \State $\inlinevec{m_x, m_y, m_z}[t_{1:N}] \gets \left(\quat{\inertiali}{\earthi}[t_{1:N}]\quatmult\quat{\imui}{\inertiali}[t_{1:N}]\right)\quatmult\nvec{m}{}[t_{1:N}]\quatmult\left(\quat{\inertiali}{\earthi}[t_{1:N}]\quatmult\quat{\imui}{\inertiali}[t_{1:N}]\right)\quatinv$
        \State $\delta_\mathrm{mag}[t_{1:N}] \gets \atantwo(m_x[t_{1:N}], m_y[t_{1:N}])$
        \State $\delta_i[t_{1:N}] \gets$ run heading correction filter with magnetic disturbance rejection on $\delta_\mathrm{mag}[t_{1:N}]$
        \State $\delta_i[t_{1:N}] \gets$ run heading correction filter with magnetic disturbance rejection on $\delta_i[t_{N:1}]$
        \State \textbf{return} $\quat{\inertiali}{\earthi}[t_{1:N}]\quatmult\quat{\imui}{\inertiali}[t_{1:N}]$,\Comment{6D sensor orientation \quat{\imui}{\earthi}}
        \State \phantom{\textbf{return}} $\inlinevec{\cos\frac{\delta_i[t_{1:N}]}{2},0,0,\sin\frac{\delta_i[t_{1:N}]}{2}}\quatmult\quat{\inertiali}{\earthi}[t_{1:N}]\quatmult\quat{\imui}{\inertiali}[t_{1:N}]$\Comment{9D sensor orientation \quat{\imui}{\earth}}
    \EndProcedure
\end{algorithmic}
{\footnotesize
VQF: real-time implementation, returns magnetic disturbance state, bias estimate, and bias estimation covariance\\
integrateGyr: gyroscope strapdown integration by \autoref{eq:gyrstrapdown}\\
$\operatorname{filtfiltLPF}$: forward-backward filtering with second-order Butterworth low-pass filter
}
\end{algorithm*}

To derive this, consider a simplified case of a Kalman filter for a system with one constant state ($x_k=x_{k-1}$) and direct measurement of the state ($C=1$):
\begin{align}
p_k^- &= p_{k-1}^+ + v\\
k_k &= \frac{p_k^-}{w + p_k^-}\\
\hat{x}_k &= \hat{x}_{k-1} + k_k (y_k - \hat{x}_{k-1})\\
p_k^+ &= p_k^- - k_k p_k^-\label{eq:simpleKFlast}.
\end{align}
In the converged state, $k_k=k_{k-1}$ and $p_k^+=p_{k-1}^+$.
From \autoref{eq:simpleKFlast} follows
\begin{align}
p_k^+ &= p_k^+ + v - \frac{p_k^+ + v}{p_k^+ + v + w} (p_k^+ + v)\\
v &= \frac{(p_k^+ + v)^2}{p_k^+ + v + w}\\
w &= \frac{(p_k^+ + v)^2}{v} - p_k^+ - v = \frac{(p_k^+)^2}{v} + p_k^+.
\end{align}
The relation given in \autoref{eq:vqf_measurement_noise_from_sigma} is then obtained by replacing the variance $p_k$ with $\sigma^2$.

Note that, for the 3-dimensional bias estimate, the $3\times3$ covariance matrix might not be close to a diagonal matrix, especially if the same sensor axis is vertical for a long time.
The uncertainty $\sigma$ of the bias estimate (in the worst-case direction) can be derived from the largest eigenvalue of the covariance matrix $\mathbf{P}$.
To avoid calculating eigenvalues, the Gershgorin circle theorem can be leveraged to obtain an upper bound estimate via the largest absolute row sum of $\mathbf{P}$.

\section{Magnetic Disturbance Rejection Algorithm}\label{app:vqf_mag_dist_rejection}

The full algorithm for the magnetic disturbance detection and rejection extension as described in \autoref{sec:vqf_mag_dist_rejection} is given in \autoref{alg:magdist}.

\begin{table*}[H]
\small
\caption{Results of \tagpx-based parameter tuning for all IOE algorithms used in the evaluation}\label{tab:vqf_algorithm_params}
\begin{tabular}{lrlrl}
\toprule
\textbf{Algorithm} & \textbf{\tagpx{}} & \textbf{Param.} & \textbf{Value} & \textbf{Search grid} {\footnotesize (start\,:\,step\,:\,end)}\\\midrule
\textbf{VQF} & \ang{2.64} & \tauAcc & 3 & 1\,:\,0.5\,:\,10\\
& & \tauMag & 9 & 1\,:\,1\,:\,30\\\midrule
MAH \cite{mahony2008nonlinear} & \ang{14.83} & $K_P$ & 1.44 & 0.02\,:\,0.02\,:\,4\\
& & $K_I$ & 0.0027 & 0\,:\,0.0001\,:\,0.004\\\midrule
MAD \cite{madgwick2010efficient} & \ang{12.01} & $\beta$ & 0.29 & 0.01\,:\,0.01\,:\,1\\
& & $\zeta_\mathrm{bias}$ & 0 & 0\,:\,0.00001\,:\,0.001\\\midrule
VAC \cite{valenti2015keeping} & \ang{5.63} & $\alpha_\mathrm{acc}$ & 0.00085 & 0.0001\,:\,0.00005\,:\,0.001\\
& & $\beta_\mathrm{mag}$ & 0.0005 & 0.0001\,:\,0.00005\,:\,0.001\\
& & $\alpha_\mathrm{bias}$ & 0.00055 & 0.0001\,:\,0.00005\,:\,0.001\\
& & $b_\mathrm{est}$ & true & \{false, true\}\\
& & $\alpha_\mathrm{adapt}$ & false & \{false, true\}\\\midrule
FKF \cite{guo2017novel} & \ang{9.19} & $\sigma^2_\mathrm{gyr}$ & 0.001 & 0.001\,:\,0.001\,:\,0.001\\
& & $\sigma^2_\mathrm{acc}$ & 0.002 & 0.001\,:\,0.0001\,:\,0.005\\
& & $\sigma^2_\mathrm{mag}$ & 0.0033 & 0.001\,:\,0.0001\,:\,0.005\\\midrule
SEL \cite{seel2017eliminating} & \ang{4.58} & \tauAcc & 3.2 & 1\,:\,0.2\,:\,5\\
& & \tauMag & 10 & 1\,:\,1\,:\,20\\
& & $\zeta_\mathrm{bias}$ & 5 & 0\,:\,1\,:\,10\\
& & $r_\mathrm{acc}$ & 2 & 0\,:\,1\,:\,10\\\midrule
MKF  & \ang{7.58} & $\sigma^2_\mathrm{acc}$ & 0.00028171 & \multirow{2}{13em}{\footnotesize \textcolor{gray}{MKF parameters were iteratively determined with line searches instead of a grid search.}}\\
& & $\sigma^2_\mathrm{mag}$ & 14.55188 &\\
& & $\sigma^2_\mathrm{gyr}$ & 0.15625 &\\
& & $\sigma^2_\mathrm{gyrdrift}$ & \num[scientific-notation=true]{0.000000000000000000003} &\\
& & $\sigma^2_\mathrm{linacc}$ & 0.49128 &\\
& & $d_\mathrm{linacc}$ & 0.81297 &\\
& & $\sigma^2_\mathrm{magdist}$ & 0.12329 &\\
& & $d_\mathrm{magdist}$ & 0.51005 &\\\midrule
KOK \cite{kok2019fast} & \ang{11.70} & $\sigma_\mathrm{gyr}$ & 0.185 & 0.01\,:\,0.005\,:\,0.5\\
& & $\zeta_\mathrm{bias}$ & 0 & 0\,:\,0.00001\,:\,0.001\\
& & $m_\mathrm{est}$ & true & \{false, true\}\\\midrule
RIANN \cite{weber2021riann}& \textcolor{gray}{\ang{1.32}}& \multicolumn{3}{l}{\textcolor{gray}{no parameters}}
\\& \multicolumn{4}{p{23em}}{\footnotesize \textcolor{gray}{\tagpx{} value not comparable to other algorithms because the inclination RMSE is used instead of the orientation RMSE.}}\\\bottomrule
\end{tabular}
\end{table*}

\section{Offline Orientation Estimation Algorithm}\label{app:vqf_offline}

The full algorithm for the acausal offline algorithm variant OfflineVQF as described in \autoref{sec:vqf_offline} is given in \autoref{alg:vqf_offline}.

\newpage

\section{Parameter Tuning for the Evaluated IOE Algorithms}\label{app:vqf_evaluation_parameter_tuning}

To provide a fair comparison for all evaluated algorithms, all algorithm parameters were tuned to provide the smallest possible \tagpx{} values.
To determine the parameters, a grid search was performed, i.e., the algorithm performance was evaluated on a grid defined by the Cartesian product of the linearly spaced parameter sets presented in \autoref{tab:vqf_algorithm_params}.
This search grid was iteratively adjusted to ensure that the distance between parameter values is sufficiently small and that the \tagpx{} parameters do not lie at the border of the grid.
The resulting averaged error, the associated parameters, and the parameter search range are presented in \autoref{tab:vqf_algorithm_params}.

Due to the high dimensionality of the search space and the slow implementation, the parameters for MKF were only evaluated using a line search, i.e., only one parameter was changed while the other parameters are kept at the previously found minimum.
The search range was also iteratively adjusted until it converged to a stable minimum.

\end{document}